\documentclass[aps,prb,preprintgroupedaddress,eqsecnum,draft,showpacs,intlimits,amsmath,amssymb]{revtex4}
\usepackage{bm}
\usepackage[final]{graphicx}
\usepackage{longtable}
\newcommand{\e}{\mathrm{e}}

\newcommand{\be}{\begin{equation}}
\newcommand{\ee}{\end{equation}}
\newcommand{\bea}{\begin{eqnarray}}
\newcommand{\eea}{\end{eqnarray}}

\newcommand{\eqan}[1]{\begin{eqnarray*}#1\end{eqnarray}}

\newcommand{\mli}[1]{\begin{multline}#1\end{multline}}

\newcommand{\ali}[1]{\begin{align}#1\end{align}}

\newcommand{\eq}[1]{\begin{equation}#1\end{equation}}

\newcommand{\lne}[1]{{\rm ln}\,{#1}}

\renewcommand{\c}[1]{\ensuremath{c_{#1_{0}}}}
\renewcommand{\k}{\ensuremath{\kappa}}
\newcommand{\Ei}[1]{{\rm Ei}({#1})}

\begin{document}

\title{Polymer chains in confined geometries: Massive field theory approach.}
\author{D.Romeis}
\affiliation{Leibniz Institute for Polymer Research Dresden e.V.,
01069 Dresden, Germany}
\author{Z.Usatenko}
 \affiliation{Institute for Condensed Matter
Physics, National Academy of Sciences of Ukraine, 79011 Lviv,
Ukraine}

\vspace{0.1cm}
\date{\today}

\begin{abstract}

The massive field theory approach in fixed space dimensions $d=3$ is
applied to investigate a dilute solution of long-flexible polymer
chains in a good solvent between two parallel repulsive walls, two
inert walls and for the mixed case of one inert and one repulsive
wall. The well known correspondence between the field theoretical
$\phi^4$ $O(n)$-vector model in the limit $n\to 0$ and the behavior
of long-flexible polymer chains in a good solvent is used to
calculate the depletion interaction potential and the depletion
force up to one-loop order. Our investigations include modification
of renormalization scheme for the case of two inert walls. The
obtained results confirm that the depletion interaction potential
and the resulting depletion force between two repulsive walls are
weaker for chains with excluded volume interaction (EVI) than for
ideal chains, because the EVI effectively reduces the depletion
effect near the walls. Our results are in qualitative agreement with
previous theoretical investigations, experimental results and with
results of Monte Carlo simulations.

\end{abstract}
\vspace{0.2cm} 

\maketitle

\renewcommand{\theequation}{\arabic{section}.\arabic{equation}}
\section{Introduction}
\setcounter{equation}{0}

Solutions of long flexible polymer chains in confined geometries
such as thin films, porous media or mesoscopic particles dissolved
in the solution have been extensively studied during last years,
including experimental, numerical and theoretical investigations.
These investigations showed that polymer solutions and binary liquid
mixture in confined geometries gave rise to a new phenomena not
observed in the bulk. The confinement of critical fluctuations of
the order parameter in a binary liquid mixture leads to an effective
long-ranged forces between the confining walls or particles immersed
in fluid as it was predicted by \cite{FdeG78}. It is named critical
(or thermodynamic) Casimir force. Such fluctuation-induced forces
are omnipresent in the nature. For example, such forces arise from
the confinement of quantum fluctuations of the electromagnetic field
and due to the well known quantum-electrodynamic Casimir effect
\cite{C48}. In polymer solutions the reason for this depletion force
originates from the presence of depletion zones near the confining
walls or mesoscopic particles due to the additional amount of
entropic energy for polymers confined within the slit or between
colloidal particles. For entropic reasons the polymer chains avoid
the space between the walls or two close particles. This leads to an
unbalanced pressure from outside which pushes the two walls or two
colloidal particles towards each other. In the case of addition of
the polymer chains to the solvent of colloidal solution effective
attraction between particles leads to flocculation \cite{SHT80}.
Such solvent-mediated flocculation mechanism was observed
experimentally for silica spheres immersed in the binary liquid
mixture of water and 2,6-lutidine \cite{BE85,GKM92,KM95}. Improving
of the experimental technique allowed recently even measure with
high accuracy the depletion force between a wall and a single
colloidal particle \cite{OSO97,RBL98,VCLY98,HHGDB08}. It should be
mentioned, that the case of two parallel walls gives the possibility
via the Derjaguin approximation \cite{D34} to describe the case of
big colloidal spherical particle, whose radius $R$ is large than the
radius of gyration $R_{g}$ and the distance between particle and the
wall $L$. It indicates, that the investigation of the case of
polymer solutions confined to geometry of two parallel walls is
important not only for description of polymer solutions confined to
film geometry and porous media, but it is also interesting from the
point of view of investigation of behavior of big colloidal
particles in polymer solutions.

During long period the interaction between polymers and colloidal
particles has been modeled by approximating the polymer chains as
hard spheres \cite{AO54,AO58}. But, such approach does not give
possibility to describe correctly behavior of small colloidal
particles in polymer solution and for the case of colloidal particle
of the big size the difference between theoretical predictions and
experimental data are bigger than $10{\%}$. In accordance with this
more effective were approaches which took into account the chain
flexibility. For example, for the case of strongly overlapping
polymer chains as it has place for the case of semidilute solution,
the chain flexibility is taken into account via phenomenological
scaling theory \cite{JLdeG79,deGS79} or self-consistent field theory
\cite{O96}. In the case of dilute polymer solution different polymer
chains do not overlap and the behavior of such polymer solution can
be described by a single polymer chain using the model of random
walk (for an ideal chain at $\theta$-solvent) or self-avoiding walk
(the real polymer chain with excluded volume interaction). The last
case corresponds to the situation when solvent temperature is above
the $\theta$-point (good solvent) and polymer coils are less compact
than in the case of ideal chains. The remarkable progress in the
investigation of this task was achieved by \cite{E97,SHKD01} via
using of dimensionally regularized continuum version of the field
theory with minimal subtraction of poles in $\epsilon=4-d$, where
$d$ is dimensionality of the space. But, as it is easy to see
\cite{SHKD01}, still there are a lot of unsolved problems and the
question arises : "How to find the theory which allows to explain
experimental data in a better way?".
 One of the methods, which up to our knowledge has not yet been applied to
this task is the massive field theory approach. This method, as it
was shown for the case of infinite \cite{Par80,Parisi},
semi-infinite \cite{DSh98} systems and specially for the case of
dilute polymer solutions in semi-infinite geometry \cite{U06} gives
better agreement with experimental data and results of the Monte
Carlo calculations. In accordance with this, the emphasis of the
present work is on the investigation of dilute polymer solution
confined to geometry of two parallel walls using the massive field
theory approach in fixed dimension $d=3$.

The most remarkable properties of fluctuation-induced forces is
their universality. They are independent of most microscopic details
and depend only on a few macroscopic properties such as the
adsorption properties of the confining walls or shape of the
particles. In accordance with this we used different combinations of
confining walls, i.e. we performed calculations for the case of two
repulsive walls, two inert walls and mixed case of one repulsive and
one inert wall. Besides, taking into account the Derjaguin
approximation \cite{D34} we obtained results for colloidal particles
of big radius near the wall and compare the obtained results with
experimental data\cite{RBL98}. In the case of two repulsive walls we
found good agreement of our results with results of Monte Carlo
simulations \cite{MB98,HG04}.

\section{The Model}
We shall assume that the solution of polymer chains is sufficiently
 dilute, so that interchain interactions and overlapping between
 different chains can be neglected, and it is sufficient to consider
 the configurations of a single chain. Long flexible polymer chains in a good
 solvent are perfectly described by a model of self-avoiding walks (SAW) on a regular
lattice \cite{CJ90}, \cite{Sh98}. Taking into account the
polymer-magnet analogy developed by \cite{deGennes}, their scaling
properties in the limit of an infinite number of steps $N$ may be
derived by a formal $n \to 0$ limit of the field theoretical
$\phi^4$ $O(n)$- vector model at its critical point.  The average
square end-to- end distance, the number of configurations with one
end fixed and with both ends fixed at the distance $x=\sqrt{({\vec
x}_{A}-{\vec x}_{B})^{2}}$ exhibit the following asymptotic behavior
in the limit $N\to \infty$ \be <R^2>\sim N^{2\nu},\quad\quad\quad
Z_{N}\sim q^{N}N^{\gamma-1},\quad\quad\quad Z_{N}(x)\sim
q^{N}N^{-(2-\alpha)},\label{RZ} \ee
 respectively.  $\nu$, $\gamma$ and $\alpha$ are the universal
 correlation length, susceptibility and specific heat critical
 exponents for the $O(n)$ vector model in the limit $n\to 0$, $d$ is the space dimensionality, $q$
 is a non universal fugacity. $1/N$ plays a role of a critical
 parameter  analogous to the reduced  critical temperature in
 magnetic systems.

 In the case when the polymer solution is in contact with a
 solid substrate, then the monomers interact with the
 surface. At temperatures, $T<T_a$, the
 attraction between the monomers and the surface leads to a critical adsorbed
 state, where a finite fraction of the monomers is attached to the
 wall and form $d-1$ dimensional structure. Deviation from the adsorption
 threshold $(c\propto(T-T_a)/T_a)$ changes sign at the transition between the
adsorbed (so-named normal transition, $c<0$) and the nonadsorbed
state (ordinary transition, $c>0$) and it plays a role of a
 second critical parameter. The adsorption threshold for long-flexible
 infinite polymer chains, where $1/N\to 0$ and $c\to 0$ is a multicritical phenomenon.

The aim of the present investigations is to describe the behavior of
such dilute solution of long-flexible polymer chains confined to a
slit geometry of two parallel walls located at the distance $L$ one
from another in $z$- direction such that the surface of the bottom
wall is located at $z=0$ and the surface of the upper wall is
located at $z=L$. Each of the two surfaces of the system is
characterized by a certain surface enhancement constant $c_{i}$,
where $i=1,2$. The correspondent effective Landau-Ginzburg
Hamiltonian describing such system is:

\bea {\cal H}_{||}[{\vec \phi}] &=&\int d^{d-1}r \int_{0}^{L} d
z\bigg\lbrace \frac{1}{2} \left( \nabla{\vec{\phi}} \right)^{2}
+\frac{1}{2} {\mu_{0}}^{2} {\vec{\phi}}^{2} +\frac{1}{4!} v_{0}
\left({\vec{\phi}}^2 \right)^{2}
\bigg\rbrace\nonumber\\
&+&\frac{c_{1_0}}{2} \int d^{d-1}r {\vec{\phi}}^{2}({\bf r},z=0)
+\frac{c_{2_0}}{2}\int d^{d-1}r {\vec{\phi}}^2({\bf r},z=L),
\label{hamiltonianslit}\eea

where ${\vec \phi}({\bf x})$ is an $n$-vector field with the
components $\phi_i(x)$, $i=1,...,n$ and  ${\bf{x}}=({\bf r},z)$,
$\mu_0$ is the "bare mass", $v_0$ is the bare coupling constant
which characterizes the strength of the excluded volume interaction
(EVI). The surfaces introduce an anisotropy into the problem, and
directions parallel and perpendicular to the surfaces are no longer
equivalent. In accordance with the fact that we have to deal with
the slit geometry $({\bf x}=({\bf r},0\leq z\leq L))$, only parallel
to surfaces Fourier transforms in $d-1$ dimensions take place. The
interaction of the polymer chain with the walls is implemented by
the different boundary conditions. As it was mentioned above, we
consider the case of two repulsive walls (the Dirichlet-Dirichlet
boundary conditions) \be c_{1}\to +{\infty},\quad c_{2}\to
+{\infty}\quad or\quad{\vec \phi}({\bf {r}},0)={\vec \phi}({\bf
{r}},L)=0\label{DD}, \ee
 two inert walls (the Neumann-Neumann boundary conditions)
\be c_{1}=0,\quad c_{2}=0\quad or\quad\frac{\partial{\vec \phi}({\bf
{r}},z)}{\partial z}|_{z=0}=\frac{\partial{{\vec \phi}({\bf
{r}},z)}}{\partial z}|_{z=L}=0\label{NN},\ee
 and the mixed case of one repulsive and one
inert wall ( the Dirichlet-Neumann boundary conditions) \be c_{1}\to
+{\infty},\quad c_{2}=0\quad or\quad {\vec \phi}({\bf {r}},0)=0,
\quad\frac{\partial{{\vec \phi}({\bf {r}},z)}}{\partial
z}|_{z=L}=0\label{DN}.\ee

The requirement in Eq.(\ref{NN}) describing the inert character of
the walls corresponds to the fixed point of the so-named special
transition \cite{DD81,D86,DSh98} in field theoretical treatment.

 In the present case the only relevant lengths are the average
 end-to-end distance $\xi_R=\sqrt{<R^2>}\sim N^{\nu}$ and the
 length $L$ -- the distance between two walls. The properties of the
 system depend on the ratio $L/\xi_R$. It should be mentioned, that
 the present field-theoretical approach is not able to describe the
 dimensional crossover from $d$ to $d-1$-dimensional systems which arises for $L<<\xi_{R}$.
 In this case the system is characterized by another
 critical temperature (see, for example, on situation in magnetic or liquid thin films)
 and moves to a new critical fixed point.

  In accordance with this the present theory is valid for the case
  $L>>\xi_{R}$. Nevertheless, we performed some assumptions, which
  allowed us to describe the region $L<<\xi_{R}$.

The well-known arguments of the polymer-magnet analogy
\cite{deGennes,CJ90,Sh98,E93} assume the correspondence between the
partition function $Z_{\parallel} ({\bf x},{\bf x}')$ of polymer
chain with ends fixed at ${\bf x}$ and ${\bf x}'$ immersed in the
volume containing the two parallel walls and the two-point
correlation function $<{\vec\phi({\bf x})}{\vec{\phi}}({\bf x}')>$
in the field theoretical $\phi^4$ $O(n)$- vector model at the formal
limit $n\to 0$ in the restricted geometry:

\be Z_{\parallel}({\bf x},{\bf x}';N,L,v_{0})={\cal
IL}_{\mu_{0}^2\to N}(<{\vec \phi}_1({\bf x}){\vec\phi}_1({\bf
x'})>|_{n=0})\label{critpoly} \ee

Here the r.h.s. denotes the Inverse Laplace transform $\mu^2\to N$
of the two point correlation function for a system modelled via the
corresponding Landau-Ginzburg Hamiltonian in the limit, where the
number of components $n$ tends to zero. $N$ determines the number of
monomers of the polymer chain and represents only an auxiliary
parameter, the trace along the chain and fixes its {\it size}
globally. The most common parameter in polymer physics to denote the
size of a polymer chains which can be observable in experiments is
$R_g$ \cite{CJ90},\cite{Sh98},\cite{E93}):
 \be
R_g^2\,=\chi^2_d\frac{R_x^2}{2}\label{RxRg}, \ee where $\chi_d$ is a
universal numerical prefactor which depends on the dimension $d$ of
the system. For an ideal polymer chains one has
$\chi^2_d=\frac{d}{3}$ and for three dimensional case $N$ equals
$R_{x}^2/2$. For the chains with EVI it could be obtained within a
perturbation expansion \cite{CJ90}.

The basic element in our calculations is the Gaussian two-point
correlation function (or the free propagator) $<{\vec \phi}_{i}({\bf
x}){\vec \phi}_{j}({\bf x}')>_{0} $ in the mixed ${\bf p}z$
representation of the form

\bea {\tilde G}_{\parallel}({\bf
p},z,z';\mu_{0},c_{1_{0}},c_{2_{0}},L) =  \frac{1}{2 \kappa_0}
((\kappa_0^2+\kappa_0(c_{1_{0}}+c_{2_{0}})+c_{1_{0}}c_{2_{0}})e^{\kappa_0
L}
&-&(\kappa_0^2-\kappa_0(c_{1_{0}}+c_{2_{0}})+c_{1_{0}}c_{2_{0}})e^{-\kappa_0
L})^{-1}\nonumber\\
((\kappa_0^2+\kappa_0(c_{1_{0}}+c_{2_{0}})+c_{1_{0}}c_{2_{0}})e^{\kappa_0(L-|z-z'|)}
&+&(\kappa_0^2
-\kappa_0(c_{1_{0}}+c_{2_{0}})+c_{1_{0}}c_{2_{0}})e^{-\kappa_0(L-|z-z'|)}\nonumber\\
+(\kappa_0^2+\kappa_0(c_{2_{0}}-c_{1_{0}})-c_{1_{0}}c_{2_{0}})e^{\kappa_0(L-z-z')}
&+&(\kappa_0^2-\kappa_0(c_{2_{0}}-c_{1_{0}})-c_{1_{0}}c_{2_{0}})e^{-\kappa_0(L-z-z')}),\nonumber\\
\label{g0slit} \eea with $\kappa_{0}=\sqrt{p^2+\mu^2_{0}}$, where
${\bf p}$ is $d-1$ dimensional moment. It should be mentioned, that
in the case $L\to\infty$  and $0\leq z,z'<<L$ (or $0<< z, z'\leq L$)
the free propagator (\ref{g0slit}) reproduces the free propagator of
the semi-infinite model (see \cite{DSh98}).

\section{Thermodynamical description}
We consider a dilute solution of long-flexible polymer chains with
the slit and allow of the polymer coils exchange between the slit
and a reservoir outside the slit. Thus the polymer solution in the
slit is in equilibrium contact with an equivalent solution in the
reservoir. We follow the thermodynamical description of the problem
as given in \cite{SHKD01}. The free energy of interaction between
the walls in such a grand canonical ensemble is defined as the
difference of the free energy of an ensemble where the wall
separation is fixed at finite distance $L$ and that where the walls
are separated on infinite distance one from another:

\be \delta F = -k_B T \,{\cal N}\,\ln\left(\frac{{\cal Z}_{||}
(L)}{{\cal Z}_{\parallel}(L\to\infty)}\right)\,,\label{dF2}\ee
 where $\cal{N}$ is the total amount of polymers in the solution and $T$ is
the temperature. ${\cal Z}_{\parallel}(L)$ is the partition function
of a polymer chain located in volume $V$ containing the walls at a
distance $L$:

\be {\cal Z}_{\parallel} (L)=\int_V\int_V\,d^dx\,d^dx'\,{\cal
Z}_{\parallel}({\bf x},{\bf x}')~,\label{outintZ}\ee

with ${\cal Z}_{\parallel}({\bf x},{\bf x}')$ representing the
partition function of a single polymer chain in the slit with its
ends fixed at points ${\bf x}$ and ${\bf x}'$.  For convenience we
can renormalise the partition functions ${\cal Z}_{\parallel}(L)$
and ${\cal Z}_{\parallel}(L\to\infty)$ on the partition function $Z$
of one polymer chain in the volume $V$ without the walls. The volume
of system $V$ can be divided into two independent subsystems $V_i$
and $V_o$ which correspond to the volume inside and outside the
slit, respectively. This gives possibility to split the term $\ln
(\frac{{\cal Z}_{\parallel}(L)}{\cal Z})$ into two parts \be
\frac{1}{V}\int_{V_{o}}d^{d}x(\frac{\hat{\cal Z}_{o}(z)}{\hat{\cal
Z}_{b}}-1)+\frac{1}{V}\int_{V_{i}}d^{d}x(\frac{\hat{\cal
Z}_{i}(z)}{\hat{\cal Z}_{b}}-1),\label{VOVI} \ee with ${\cal Z}=V
\hat{\cal Z}_{b}$, $\hat{\cal Z}_{b}=\int_{V}d^{d}x' {\cal
Z}_{b}({\bf x},{\bf x}')$ where ${\cal Z}_{b}({\bf x},{\bf x}')$ is
the partition function of one polymer chain in the unbounded
solution with fixed ends at ${\bf x}$ and ${\bf x}'$ and $\hat{\cal
Z}_{o,i}(z)=\int_{V_{o,i}}d^{d}x' {\cal Z}_{\parallel}({\bf x},{\bf
x}')$.

 In the thermodynamical limit (as ${\cal N},V\to\infty$) the contribution
 from the first term in (\ref{VOVI}) disappear and the reduced free energy of interaction $\delta f$ per
 unit area $A=1$ of the confining walls may be written as:
 \bea \delta
f =\frac{\delta F}{n_p k_B T} = L
&-& \int_{V_i} d^dx \frac{{\hat{\cal Z}}_i (z)}{{\hat{\cal Z}}_b}\\
+\int_{V_{HS_{1}}}d^{d}x \left(\frac{{\hat{\cal
Z}}_{HS_1}(z)}{{\hat{\cal Z}}_b}-1\right)
&+&\int_{V_{HS_{2}}}d^{d}x\left(\frac{{\hat{\cal
Z}}_{HS_2}(z)}{{\hat{\cal Z}}_b}-1\right),\label{df}\eea

where $n_p = {\cal N}/V$ is the number density of the polymer chains
in the bulk solution and
 \eq{ \label{oneendint2} {\hat{\cal
Z}}_{HS_i}(z) =  \int_{V_{HS}} d^d x' {\cal Z}_{HS_i}({\bf x},{\bf
x'})\,, } with $i=1,2$ and ${\cal Z}_{HS_i}({\bf x},{\bf x'})$
denoting the corresponding partition functions for a polymer chain
in a half space with two fixed ends at points ${\bf x}$ and ${\bf
x'}$. The functions ${\hat{\cal Z}}_i (z)$ and ${\hat{\cal
Z}}_{HS_i} (z)$ depend only on the $z$-coordinates perpendicular to
the walls. The {\it reduced free energy of interaction} $\delta f$,
according to (\ref{df}), is a function of the dimension of a length
and dividing it by another relevant length scale (namely that for
the size of the chain in the bulk, e.g. $R_x$) yields a universal
dimensionless scaling function \eq{ \label{Theta} \Theta(y) =
~\frac{\delta f}{R_x}\,, } where $ y = L/R_x $ is a dimensionless
scaling variable. The resulting depletion force between the two
walls induced by the polymer solution is denoted as: \eq{
\label{Gamma} \Gamma(y) =~-\frac{d (\delta
f)}{dL}~=~-\frac{d\Theta(y)}{dy}~. \vspace*{0.5cm}}

The total grand canonical free energy $\Omega$ of the polymer
solution with the slit is: \eq{ \label{tdlimitgrandcan} \Omega =
-n_p\,k_b\,T\,A\,L\omega }  with  \eq{ \label{omega}
\omega\,=\frac{1}{L}\int_0^L\,dz\,\frac{\hat{\cal Z}_i(z)}{\hat{\cal
Z}_b}. } Taking into account (\ref{df}) and (\ref{tdlimitgrandcan})
we can write for unit surface area $A=1$: \be
\frac{\Omega}{n_{p}k_{B}T}=f_b\,L\,+\,f_{s_1}\,+\,f_{s_2}\,+\,\delta
f\,,\label{decompgrandcan}\ee with the reduced bulk free energy per
unit volume $ f_b =-1$ and the reduced surface free energy per unit
area \be
 f_{s_i}=\int_{V_{HS_{i}}}dz\left( 1-\frac{{\hat{\cal
Z}}_{HS_i}(z)}{{\hat{\cal Z}}_b}\right).\label{surfm}\ee

Further for convenience we can introduce ${\cal X}$, the total
system susceptibility in the form
 \be
 {\cal X} =\frac{1}{V}\int_{V}\int_{V}d^dx\,
d^dx'\,< {\vec\phi}_1({\bf x}){\vec\phi}_1({\bf x'})>.\label{chi}\ee
This definition is consistent with the bulk susceptibility for the
unbounded space given as ${\cal X}_b=\frac{1}{m^2}$ to all orders of
perturbation theory
 (e.g. \cite{Parisi}). $\hat{\cal Z}_b$ being the Inverse Laplace
transform of ${\cal X}_b$  and $\hat{\cal Z}_b=1$ to all orders as
well. Accordingly to (\ref{critpoly}) and (\ref{chi}) we can rewrite
(\ref{df}) in the form

\be \delta f = {\cal IL}_{\mu^2\to R_x^2/2}\Big\lbrace\,L\, ({\cal
X}_b-{\cal
X}_{||})\,-\,\Upsilon_{1}\,-\,\Upsilon_{2}\Big\rbrace,\label{dfcrit}
\ee where ${\cal X}_{||}$ denotes the total susceptibility for a
slit geometry and $\Upsilon_{i}$ with $i=1,2$ give two half-space
(HS) contributions such that $ f_{s_i}={\cal IL}_{\mu^2\to
R_x^2/2}\big\lbrace\Upsilon_{i}\big\rbrace\,$ (see Appendix A).

\section{Correlation functions and renormalization conditions}
Correlation functions which involve $N'$ fields $\phi({\bf{x}}_{i})$
at distinct points ${\bf{x}}_{i}(1\leq i \leq N')$ in the bulk,
$M_{1}$ fields $\phi_{1}({\bf{r}}_{j_{1}},z=0)\equiv
\phi_{s_{1}}({\bf{r}}_{j_{1}})$ at distinct points on the wall $z=0$
and $M_{2}$ fields $\phi_{2}({\bf{r}}_{j_{2}},z=L)\equiv
\phi_{s_{2}}({\bf{r}}_{j_{2}})$ at distinct points on the wall
$z=L$, and $I$ insertion of the bulk operator
$\frac{1}{2}\phi^{2}({\bf{X}}_{k})$ at points ${\bf{X}}_{k}$ with
$1\leq k \leq I$, $I_{1}$ insertions of the surface operator
$\frac{1}{2}\phi_{s_{1}}^{2}({\bf{R}}_{l_{1}})$ at points
${\bf{R}}_{l_{1}}$ with $1\leq l_{1} \leq I_{1}$ and $I_{2}$
insertions of the surface operator
$\frac{1}{2}\phi_{s_{2}}^{2}({\bf{R}}_{l_{2}})$ at points
${\bf{R}}_{l_{2}}$ with $1\leq l_{2} \leq I_{2}$, have the form
\cite{D86,DSh98}

\bea G^{(N',M_{1},M_{2},I,I_{1},I_{2})}(\{{\bf x}_{i}\},\{{\bf
r}_{j_{1}}\},\{{\bf
r}_{j_{2}}\},\{{\bf{X}}_{k}\},\{{\bf{R}}_{l_{1}}\},\{{\bf{R}}_{l_{2}}\})&=&\nonumber\\
 < \prod_{i=1}^{N'} \phi({\bf
x}_{i})\prod_{j_{1}=1}^{M_{1}}\phi_{s_{1}}({\bf
r}_{j_{1}})\prod_{j_{2}=1}^{M_{2}}\phi_{s_{2}}({\bf
r}_{j_{2}})\prod_{k=1}^{I}\frac{1}{2}\phi^{2}({\bf{X}}_{k})
\prod_{l_{1}=1}^{I_{1}}\frac{1}{2}\phi^{2}_{s_{1}}({\bf{R}}_{l_{1}})
\prod_{l_{2}=1}^{I_{2}}\frac{1}{2}\phi^{2}_{s_{2}}({\bf{R}}_{l_{2}})>~~.
\label{10} \eea

Here, the symbol $<...>$ denotes averaging with Hamiltonian
(\ref{hamiltonianslit}). The free propagator of a Gaussian chain in
slit geometry in the mixed ${\bf{p}},z$ representation has the form
(\ref{g0slit}), as was mentioned above.

Taking into account that surface fields
$\phi_{s_{i}}({\bf{r}}_{j_{i}})$ and surface operators
$\frac{1}{2}\phi_{s_{i}}^{2}({\bf{R}}_{i})$ with $i=1,2$ scale with
scaling dimensions that are different from those of their bulk
analogs $\phi({\bf{x}}_{j})$ and $\frac{1}{2}\phi^{2}({\bf{X}}_{j})$
(see \cite{DSh98}), the renormalized correlation functions involving
$N'$ bulk fields and $M_{1}$ surface fields on the wall $z=0$ and
$M_{2}$ surface fields on the wall $z=L$, $I$ bulk operators,
$I_{1}$ and $I_{2}$ surface operators can be written as \bea
 G_{R}^{(N',M_{1}M_{2},I,I_{1},I_{2})} ( ;
\mu,v,c_{1},c_{2},L)=&&\nonumber\\ Z_{\phi}^{-(N'+M_{1}+M_{2})/2}
Z_{1}^{-M_{1}/2}Z_{2}^{-M_{2}/2} Z_{\phi^2}^{I}
Z_{\phi_{s_{1}}^2}^{I_{1}}
Z_{\phi_{s_{2}}^2}^{I_{2}}G^{(N',M_{1},M_{2},I,I_{1},I_{2})} ( ;
\mu_{0},v_{0},c_{1_0},c_{2_0},L)&&,\label{12} \eea where $Z_{\phi}$,
$Z_{1}$, $Z_{2}$ and $Z_{\phi^2}$, $Z_{\phi_{s_{1}}^2}$,
$Z_{\phi_{s_{2}}^2}$ are correspondent UV-finite (for $d<4$)
renormalization factors. The typical bulk and surface short-distance
singularities of the correlation functions $G^{(N',M_{1},M_{2})}$
can be removed via mass shift $\mu_{0}^{2}=\mu^2+\delta \mu^2$ and
surface-enhancement shifts $c_{i_0}=c_{i}+\delta c_{i}$,
respectively \cite{DSh98}.
 The renormalizations
of the mass $\mu$, coupling constant $v$ and the renormalization
factor $Z_{\phi}$ are defined by standard normalization conditions
of the infinite-volume theory \cite{Parisi,BGZ76,A84,Z89,ID89}. In
order to adsorb uv singularities located in the vicinity of the
surfaces, a surface-enhancement shifts $\delta c_{i}$ are required.
In this connection the new normalization conditions should be
introduced. It is obvious, that in the limit $L\to \infty$ we should
have
 \bea \lim_{L\to\infty}\left[{\tilde
G}^{(0,2,0)}_{R}({\bf
p};\mu,v,c_{1},c_{2},L)\vert_{p=0}\right]&=&\frac{1}{\mu+c_1},\nonumber\\
\lim_{L\to\infty}\left[{\tilde G}^{(0,0,2)}_{R}({\bf
p};\mu,v,c_{1},c_{2},L)\vert_{p=0}\right]&=&\frac{1}{\mu+c_2}.\label{ccond1semi}
\eea For the renormalization factors $Z_{i}$, $Z_{\phi_{s_{i}}^2}$
where $i=1,2$ we obtain, respectively

\bea \lim_{L\to\infty}\left[\frac{\partial}{\partial p^2}{\tilde
G}^{(0,2,0)}_{R}({\bf p};\mu,v,
c_{1},c_{2},L)\big\vert_{p=0}\right]&=&
-\frac{1}{2\mu(\mu+c_1)^2},\nonumber\\
\lim_{L\to\infty}\left[\frac{\partial}{\partial p^2}{\tilde
G}^{(0,0,2)}_{R}({\bf
p};\mu,v,c_{1},c_{2},L)\big\vert_{p=0}\right]&=&
-\frac{1}{2\mu(\mu+c_2)^2},\label{Zsurfcondsemi} \eea and

\bea \lim_{L\to\infty}\left[{\tilde G}^{(0,2,0;0,1,0)}_{R}({\bf
p},{\bf P};\mu,v,c_{1},c_{2},L)\big\vert_{p,P=0}\right]&=&
\frac{1}{(\mu+c_1)^2},\nonumber\\
\lim_{L\to\infty}\left[{\tilde G}^{(0,0,2;0,0,1)}_{R}({\bf p},{\bf
P};\mu,v,c_{1},c_{2},L)\big\vert_{p,P=0}\right]&=&
\frac{1}{(\mu+c_2)^2}. \label{Z2surfcondsemi} \eea In the limit
$L\to\infty$ all these conditions yield exactly the same shifts
$\delta c_i$ and renormalization factors as in the semi-infinite
case. It is intuitively clear that in the case of two inert walls or
mixed walls situated on big, but finite distance $L$ with $L\gtrsim
R_{g}$ such that the chain is still not deformed too much from its
original size in the bulk, the shift of $c_0^{sp}\to c^{sp}$ may
depend on the presence of the other surface and hence on the size of
the slit. So, in the case of $L\gtrsim R_{g}$  (or $\mu L\gtrsim 1$)
from (\ref{g0slit}) and (\ref{ccond1semi}) we obtain new conditions

\bea \lim_{L\mu\gtrsim 1}\left[{\tilde G}^{(0,2,0)}_{R}({\bf
p};\mu,v, c_1,c_2,L)\vert_{p=0}\right]&=&
\frac{1}{\mu+c_1}\left(1+\frac{2\mu}{\mu+c_1}\frac{\mu-c_2}{\mu+c_2}\e^{-2\mu
L}+{\cal O}(\e^{-4\mu L})\right),\nonumber\\
\lim_{L\mu\gtrsim 1}\left[{\tilde G}^{(0,0,2)}_{R}({\bf p};\mu,v,
c_1,c_2,L)\vert_{p=0}\right]&=&
\frac{1}{\mu+c_2}\left(1+\frac{2\mu}{\mu+c_2}\frac{\mu-c_1}{\mu+c_1}\e^{-2\mu
L}+{\cal O}(\e^{-4\mu L})\right). \label{ccondslit} \eea

The above mentioned conditions (\ref{ccondslit}) give one-loop order
corrections to the respective surface-enhancement shifts $\delta
c_{i}$ of semi-infinite theory in the case of large, but finite wall
separation $L$ . In accordance with this for the case of mixed walls
we obtain \be \delta c_{1}^{S-O}=\delta c_{1}+\Delta^{(S-O)}
\label{cslitso}\ee with corrections of order ${\cal O}(e^{-2\mu L})$
 \be
\Delta^{(S-O)}\,=\,\frac{\mu}{4}\left(\frac{1}{\mu
L}+C_E+\,\ln8\,-\,3+\,\ln \mu L\,-\e^{4\mu L}{\rm Ei}(-4\mu
L)\right)\e^{-2\mu L}.\label{deltaos} \ee

In the case when both walls are inert, the modified surface
enhancement shifts are \be \delta c_{i}^{S-S}=\delta c_{i} +\Delta^
{(S-S)}\label{cslitss} \ee with
 \be
\Delta^{(S-S)}=-\Delta^{(S-O)}\,-\,\mu\left(\ln2\,-\,\frac{1}{2}\right)\e^{-2\mu
L}.\label{deltass} \ee The above mentioned corrections $\delta
c_{i}$ are UV singular for $d=3$ dimensions. They provide the
singular parts of the counterterms that cancel the UV singularities
of correspondent correlation functions by analogy as it took place
for semi-infinite systems (see \cite{DSh98}). The above mentioned
corrections $\Delta^{(S-O)}$ and $\Delta^{(S-S)}$ are finite in
$d\leq 4$ dimensions.

\section{Results for Gaussian Chains}

Let us consider at the beginning the Gaussian model for ideal
polymer chains ($v_0=0$). As mentioned above it corresponds to the
situation of a polymer chain under $\Theta$-solvent conditions.

 For general case of arbitrary $c_1$ and $c_2$ on the confining walls
 we obtain for the reduced free energy of interaction:
\bea \delta f=-\,{\cal IL}_{\mu^2\to
R_x^2/2}\Bigg\lbrace\frac{1}{\mu^3}
\Big[(\mu+c_1)(\mu+c_2)\e^{\mu L}\,&-&\,(\mu-c_1)(\mu-c_2)\e^{-\mu L}\Big]^{-1}\times\nonumber\\
\Big\lbrace4\,c_1c_2-\left(\mu(c_1+c_2)+2\,c_1c_2\right)\e^{\mu L}+
\left(\mu(c_1+c_2)-2\,c_1c_2\right)\e^{-\mu L}\Big\rbrace
&+&\,\frac{1}{\mu^3}\left(\frac{c_1}{\mu+c_1}\,+\,\frac{c_2}{\mu+c_2}\right)\Bigg\rbrace.
\label{dfgauss}\eea

 First, consider the case of the Dirichlet-Dirichlet (D-D) boundary conditions (\ref{DD}) on
the confining surfaces. Taking the limits $\frac{c_1}{m}\to
\infty,\frac{c_2}{m}\to \infty$ yields:

\be {\Theta}^{D,D}(y)=-4y\,{\cal {IL}}_{\tau\to(2y^2)^{-1}}
\left(\frac{1}{\tau^{3/2}}\frac{1}{1+\e^{\sqrt{\tau}}}\right)\,,
\label{thoogauss}\ee where $\tau=\mu^2L^2$ and $y=\frac{L}{R_x}$.
 The result indicates that if both $c_i$ being positive, the
depletion interaction potential is negative and hence the walls
attract each other due to the depletion zones near repulsive walls.
The inverse Laplace transform can only be performed numerically (the
plot is shown in Figure \ref{fig:theta_gamma_oo}) or may be expanded
for asymptotic values of $\sqrt{\tau}$.  The obtained results for
ideal polymer chains in slit of two repulsive walls are in agreement
with previous theoretical results obtained in Ref.\cite{SHKD01}.
But, it should be mentioned, that on plotting these functions the
authors of \cite{SHKD01} used a rescaled variable $\sqrt{2}R_x$,
which was not mentioned there.

 Now we proceed to the case of two inert walls, what corresponds to
 the Neumann-Neumann (N-N) boundary conditions (\ref{NN}).
For the free energy of interaction we obtain \be
{\Theta}^{N,N}(y)~=~0. \ee This corresponds to the fact that ideal
chains do not loose free energy inside the slit in comparison to the
free chains in unrestricted space. The entropy loss is fully
regained by the surface interactions provided by the two walls.

Taking the limits $\frac{c_1}{m}\to \infty, \,\frac{c_2}{m}\to 0$ in
accordance with (\ref{DN}) (the Dirichlet-Neumann (D-N) boundary
conditions) from (\ref{dfgauss}) we obtain: \be
{\Theta}^{D,N}(y)=-\,2y\,\,{\cal IL}_{\tau\to(2y^2)^{-1}}
\left(\frac{1}{\tau^{3/2}}\frac{1}{1+e^{2\sqrt{\tau}}}\right).
\label{thosgauss}\ee
 This result can only be evaluated
numerically and is plotted in Figure \ref{fig:theta_gamma_os}. Lets
consider different
asymptotic regions of $y$.\\
{\bf Wide slits ($y>>1$):}\hfill\\[0.1cm]
In the case  $\mu L>>1$ from (\ref{thoogauss}) we obtain for two
repulsive walls: \be {\Theta}^{D,D}(y)\,{\approx}\,
4y\bigg[\,{\rm{erfc}}\left(\frac{y}{\sqrt{2}}\right)-
\frac{1}{y}\sqrt{\frac{2}{\pi}}\exp\left(-\frac{y^2}{2}\right)\bigg]-
8y\bigg[\,{\rm{erfc}}\left(\sqrt{2}\,y\right)-\frac{1}{y\,
\sqrt{2\pi}}\exp\left(-2y^2\right)\bigg].\label{thooexp}\ee The
force (\ref{Gamma}) becomes \be \Gamma^{D,D}(y)\,{\approx}\,-4\,{\rm
erfc} \left(\frac{y}{\sqrt{2}}\right)\,+\,8\,{\rm
erfc}\left(\sqrt{2}\,y\right).\label{gaooexp} \ee
 And for one repulsive and one inert wall we have:
\be {\Theta}^{D,N}(y){\approx}4y\,
{\rm{erfc}}\left(\sqrt{2}\,y\right)\,-\,\frac{4}{\sqrt{2\pi}}\,
\exp\left(-2y^2\right),\label{thosexp}\ee which implies \be
\Gamma^{D,N}(y){\approx}-4\,{\rm{erfc}}\left(\sqrt{2}\,y\right).\label{gaosexp}
\ee These approximating functions are presented on Figures
\ref{fig:theta_gamma_oo} and \ref{fig:theta_gamma_os}, respectively.
\smallskip\\
{\bf Narrow slits ($y\ll1$)}\\[0.1cm]
In the case of narrow slit $\mu L<<1$ the asymptotic solution for
(\ref{thoogauss}) reads: \be
{\Theta}^{D,D}(y){\approx}\,-\frac{4}{\sqrt{2\pi}}\,+\,y\,.
 \label{thoonarrow}\ee and the force simply becomes
$\Gamma^{D,D}(y){\approx}-1$.

 For the depletion interaction potential (\ref{thosgauss}) we get:
\eq{ \label{thetanarrowos}
{\Theta}^{D,N}(y){\approx}\,-\frac{2}{\sqrt{2\pi}}\,+\,y~. } For the
force we have again $\Gamma^{D,N}(y){\approx}-1$.

 These results can be understood phenomenologically. In our units the
quantities ${\Theta}$ and ${\Gamma}$ are normalized to the overall
polymer density $n_p$. So, the above results simply indicate that
the force is entirely induced by the free chains surrounding the
slit or in other words by the full bulk osmotic pressure from the
outside of the slit. No chain has remained in the slit. It is
reasonable in the case of repulsive walls in the limit of narrow
slits. Unfortunately, the narrow slit regime is beyond the validity
of our approach in the presence of EVI, as mentioned above. But, the
above mentioned arguments can be used in order to obtain the leading
contributions to the depletion effect as $y\to 0$. We can state that
in the case of very narrow slits the chains would pay a very high
entropy to stay in the slit or even enter it. It is due to the fact
that the phase space containing all possible conformations is
essentially reduced by the squeezing confinement to a size
$\frac{d-1}{d}$\,times its original size (for an unconfined chain).
Therefore, the ratio of partition function of polymer chain in slit
and free chain partition function vanishes strongly as $y\to 0$,
which implies directly the function $\omega$ in (\ref{omega}).
Setting $\omega=0$ and using only the corresponding surface
contributions and the bulk contribution ($f_b=-1$) in
(\ref{decompgrandcan}) must lead to the same asymptotic limits in
the narrow slit regime.
 The advantage of this procedure is that no expansion necessary and it
 should be equally valid in the EVI-regime.

 In Figures \ref{fig:theta_gamma_oo},\ref{fig:theta_gamma_os}
 and \ref{fig:theta_gamma_ss} the depletion interaction
 potential $\Theta(y)$ and depletion force $\Gamma(y)$ are plotted for all boundary
conditions. As expected, the results for mixed walls are located in
between the results of two inert walls and those of two repulsive
walls.

\section{Results for good solvent}

In good solvent the EVI between chain monomers play a crucial role
so that the polymer coils occupy the bigger volume and are less
compact than in the case of ideal polymer chains. The influence of
EVI on the depletion functions can be obtained in the framework of
the massive field theory approach in fixed dimensions $d=3$ up to
one-loop order expansion of the two-point correlation functions
$G^{(2,0,0)}$ restricted in slit geometry (\ref{hamiltonianslit}).
The bare total susceptibility ${\cal X}_{\parallel}^{bare}$
(see(\ref{dfcrit})) for the slit geometry in accordance with
(\ref{chi}),(\ref{omega}) and (\ref{10}) is : \bea {\cal
X}^{bare}_{||}(\mu_0,v_0,\c{1},\c{2},L)=\frac{1}{L}
\int_0^L\int_0^Ldzdz'\bigg\lbrace\tilde{G}_{||}({\bf p}=0,z,z';\mu_0,\c{i},L)\nonumber\\
-\frac{n+2}{6}v_0\int_0^Ldz''\int_{\bf q}\tilde{G}_{||}({\bf
p}=0,z,z'';\mu_0,\c{i},L) \tilde{G}_{||}({\bf
q},z'',z'';\mu_0,\c{i},L) \tilde{G}_{||}({\bf
p}=0,z'',z';\mu_0,\c{i},L)\bigg\rbrace\,.\label{1Loop} \eea The two
HS contributions denoted by $\Upsilon_{i}$ (see (\ref{dfcrit}))can
be obtained in accordance with (\ref{surfm}) similarly to
(\ref{1Loop}) with the propagators of semi-infinite system.  Some
details for the calculation of these quantities for zero-loop and
one-loop order for different surface critical points of interest
(ordinary, special) are presented in the Appendix A.

\subsection{Two repulsive walls}
Lets consider first the case of D-D boundary conditions (\ref{DD})
on each of the two surfaces. In this case no surface divergences
appear in the calculation of the correlation functions and any
surface renormalization is not necessary at all. Each surface term
($f_{s_i}\,,\,i=1,2$) contributes: \eq{ \label{surfO}
f^{D}_{s}\,=\,\sqrt{\frac{2}{\pi}}\left(\,1\,-
\,\frac{\ln\frac{9}{8}}{4}\right)\,R_x\,. } After performing the
standard mass and coupling constant renormalization and additive
subtraction at zero momentum all divergent terms disappear and the
corespondent function ${\cal X}_{||\,ren}^{D,D}$  can be obtained.
In order to be concise, we do not present here the complicated form
for ${\cal X}_{||\,ren}^{D,D}$ and just discuss the limiting cases
 of wide and narrow slit regimes. \\
{\bf Wide slits ($y\gtrsim1$):}\hfill\\[0.1cm]

\begin{figure}[ht!]
\begin{center}
\includegraphics[width=8cm]{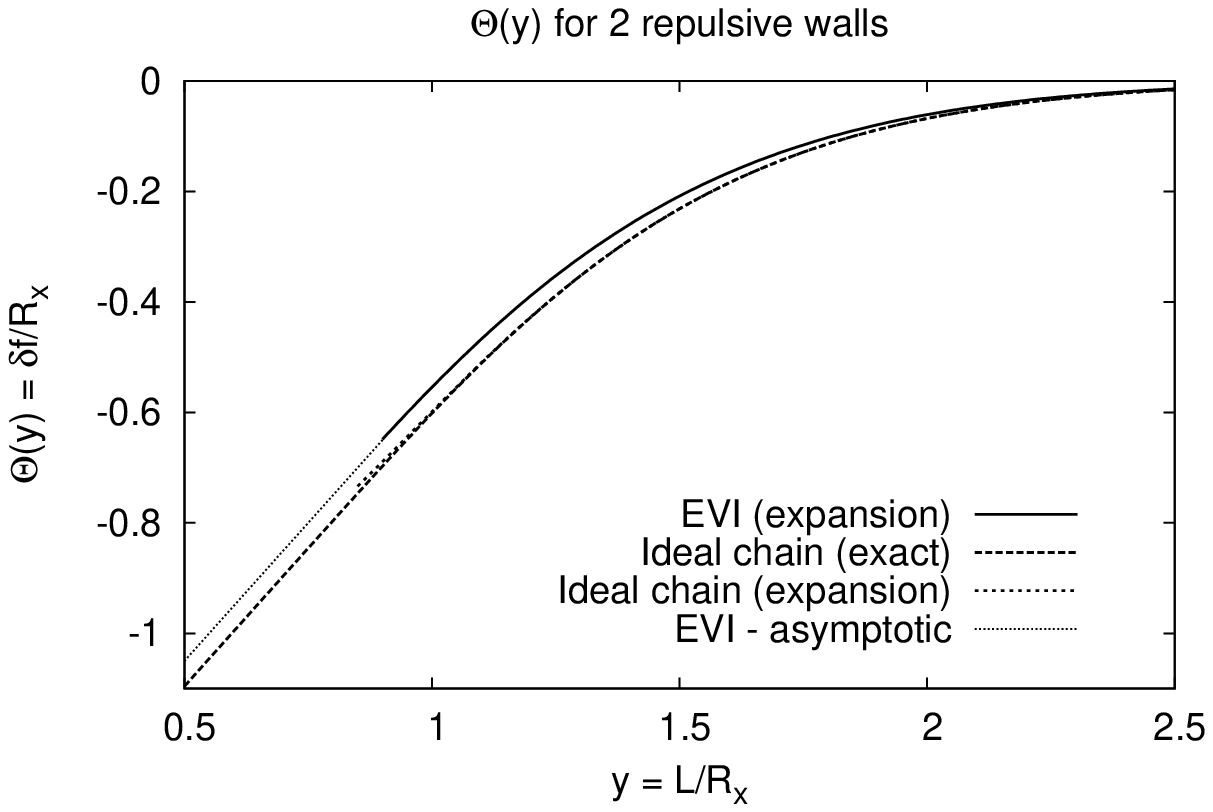}\hspace*{0.3cm}
\includegraphics[width=8cm]{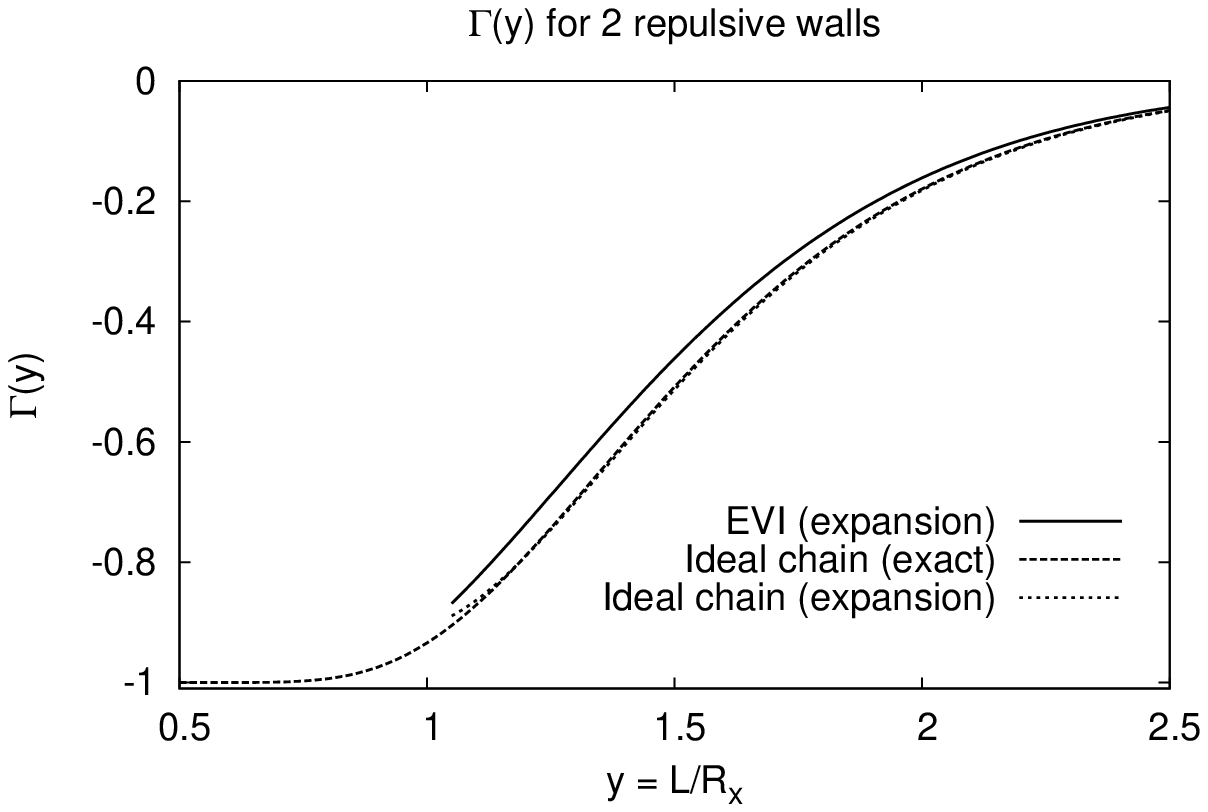}
\caption{The functions $\Theta(y)$ and $\Gamma(y)$ for two repulsive
walls} \label{fig:theta_gamma_oo}
\end{center}
\end{figure}
 The massive field-theory approach at fixed dimensions $d=3$ gives a
 rather simple result in one-loop order than results obtained in \cite{SHKD01} with help of dimensionally
regularized field theory with minimal subtraction of poles in
$\epsilon$-expansion. It should be mentioned, that in \cite{SHKD01}
a wide slit approximation was carried out as well up to the first
non trivial order (apparently ${\cal O}\left(\e^{-\mu L}\right)$).
Therefore, we performed calculations up to the next order term
$\sim{\cal O}\left(\e^{-2\mu L}\right)$. The renormalized total
susceptibility for the slit geometry up to one-loop order in $d=3$
for polymer case $n\to 0$ in the wide slits regime $\mu L>>1$ is:

\bea {\cal X}_{||\,ren}^{D,D}  L {\approx}
\frac{L}{\mu^2}\,&-&\,\frac{1}{\mu^3}\left(2-\frac{\ln\frac{9}{8}}{2}\right)\,
+\,\frac{\e^{-\mu L}}{\mu^3}\left(4-\ln\frac{3}{2}\right)\nonumber\\
-\,\frac{\e^{-2\mu L}}{\mu^3}\bigg\lbrace\frac{9-C_E-2\,
\ln\frac{3}{2}}{2}\,-\,\frac{3}{2\mu L}\,-\,\frac{\ln(\mu L)}{2}
&+&\,\e^{\mu L}\,\Ei{-\mu L}\,-\,\e^{3\mu L}\,\Ei{-3\mu
L}\,+\,\frac{\e^{4\mu L}}{2}\,\Ei{-4\mu
L}\bigg\rbrace.\label{suscoo}\eea The exponential integral
functions, denoted by $\Ei{x}$, can be expanded for large, negative
arguments as well in accordance with (see e.g. \cite{Gradshteyn}):
$\e^{x}\,\Ei{-x}=\,-\,\frac{1}{x}\,+{\cal
O}\left(\frac{1}{x^2}\right)$. Thus, for the depletion interaction
potential we obtain: \bea  \Theta(y){\approx}
\left(4-\lne{\frac{3}{2}}\right)
\left[y\,{\rm{erfc}}\left(\frac{y}{\sqrt{2}}\right)-
\sqrt{\frac{2}{\pi}}\exp\left(\frac{y^2}{2}\right)\right]
-\frac{55}{48y}\,{\rm erfc}\left(\sqrt{2}\,y\right)\nonumber\\
-\,\left(\frac{163}{24}-\lne{\frac{3}{2}}-\frac{C_E}{2}\right)\,
\left[2\,y\,\,{\rm{erfc}}\left(\sqrt{2}\,\,y\right)-\,
\sqrt{\frac{2}{\pi}}\,\,\exp\left(-2y^2\right)\right]
-\,\frac{y}{4}\,\,{\cal IL}_{\tau\to\frac{1}{2y^2}}
\left(\frac{\lne{\tau}}{\tau^{3/2}}\,\e^{-2\sqrt{\tau}}\right)~.\label{thetaoo}
\eea
 A comparison of the obtained results to the ideal chain results in a wide slit
 regime (see Figure 1) shows that the EVI
 reduces the depletion effects for two repulsive
walls.\smallskip\\
{\bf Narrow slits $(y\ll1)$:}\hfill\\[0.1cm]
Following the simple argument obtained from the discussion of the
exactly solvable ideal chain model the entire slit contribution
$\omega$ (\ref{omega}) to the reduced free energy of interaction
$\delta f$ in (\ref{decompgrandcan}) is simply set to zero and the
depletion effect is only calculated from the bulk and surface
contributions. In this limit the depletion potential becomes: \be
\Theta(y){\approx}\,y\,-\,
\frac{2\,\sqrt{2}}{\sqrt{\pi}}\left(\,1\,-
\,\frac{\ln\frac{9}{8}}{4}\right)\,.\ee and the force again is
unity. In Figure \ref{fig:theta_gamma_oo} one can follow how the two
regimes come to match in the crossover regime $y\approx1$. The
lowest order expansion in case of wide slits would not be able to
show this matching. With these two approximations we are in the
position to present a rather complete picture of the problem in
comparison to the approach given in \cite{SHKD01}.

\subsection{One repulsive / One inert wall}

This case has not been studied so far in any approach. Since we are
now dealing with an inert wall, the surface renormalization should
be taken into account. Again, the full result for the renormalized
total susceptibility in a slit system ${\cal X}_{||\,ren}$ has
complicated form and we discuss just limiting cases of wide and
narrow slits.

 The surface contribution for a repulsive wall coincides with
 (\ref{surfO}) and for inert wall we have:
 \be
f^{N}_{s}\,=\frac{2\ln2\,-\,1}{8}\sqrt{\frac{2}{\pi}}\,R_x\,\,.
\label{surfS}\ee
{\bf Wide slits ($y\gtrsim1$):}\hfill\\[0.1cm]
For the total susceptibility up to ${\cal O}\left(\e^{-2\mu
L}\right)$ order we obtain: \bea  {\cal X}_{||\,ren}^{OS}\,L
{\approx}\frac{L}{\mu^2}\,-
\,\frac{1}{\mu^3}\left(1+\frac{2\,\ln\frac{4}{3}\,-\,\frac{1}{2}}{4}\right)\,
&+&\,\frac{\e^{-\mu L}}{\mu^3}\left(\frac{2\ln4\,-\,\ln3\,-\,1}{2}\right)\nonumber\\
+\,\frac{\e^{-2\mu
L}}{\mu^3}\bigg\lbrace\frac{31-2C_E}{8}-\frac{\ln3}{2}-
\frac{7\,\ln2}{4}-\frac{3}{4\mu L}-\frac{\ln(\mu L)}{4}
&+&\frac{\e^{\mu L}}{2}\,\Ei{-\mu L}\,+\,\frac{\e^{4\mu L}}{4}\,
\Ei{-4\mu L}\,-\,\frac{\e^{3\mu L}}{2}\,\Ei{-3\mu L}\bigg\rbrace.
\label{suscos}\eea  In comparison to the result for ideal chains
(\ref{thosgauss}) where the lowest order term, contributing to the
total susceptibility in the wide slit limit is of order ${\cal
O}\left(\e^{-2\mu L}\right)$, now the additional term of order
${\cal O}(\e^{-\mu L})$ appears.
\begin{figure}[ht!]
\begin{center}
\includegraphics[width=8cm]{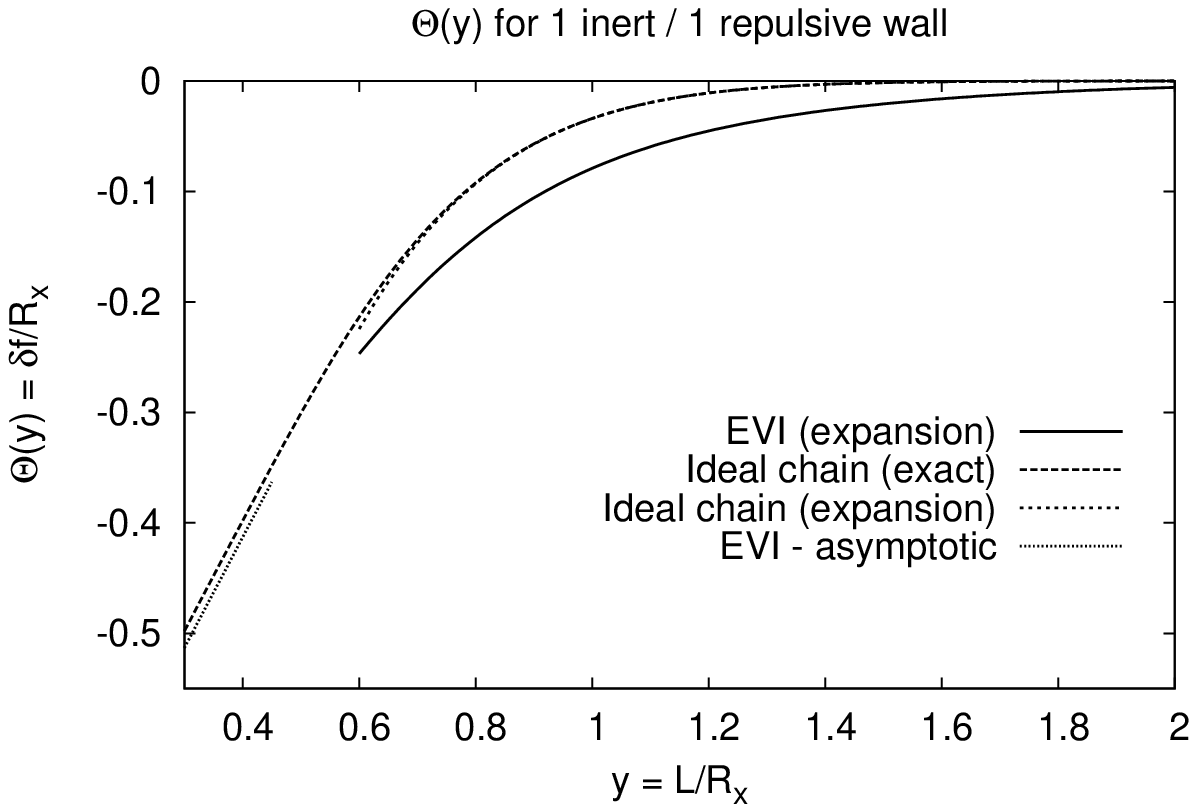}\hspace*{0.3cm}
\includegraphics[width=8cm]{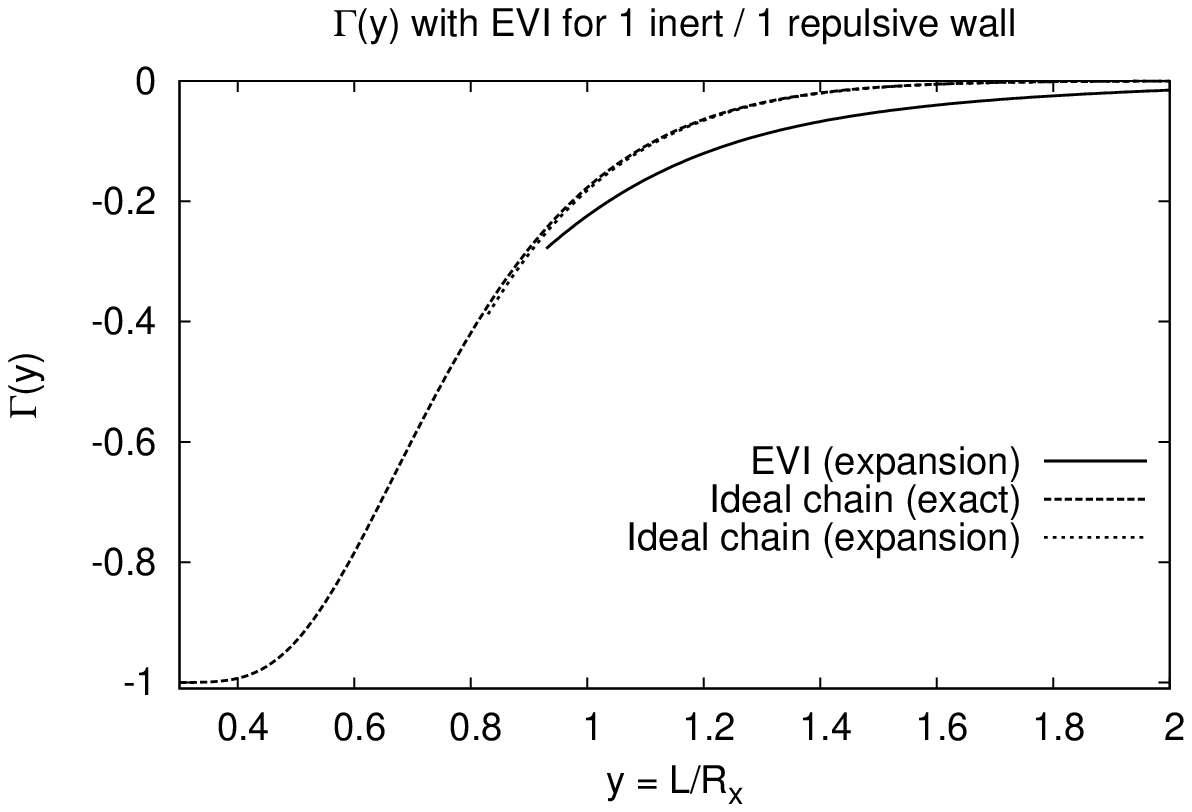}
\caption{The functions $\Theta(y)$ and $\Gamma(y)$ for one repulsive
and one inert wall with and without excluded volume interactions
(EVI)} \label{fig:theta_gamma_os}
\end{center}
\end{figure}

 In this case the depletion interaction potential becomes :
 \bea \Theta(y){\approx}\frac{1}{2}
\left(\ln{\frac{16}{3}-1}\right)\left(y\,{\rm erfc}
\left(\frac{y}{\sqrt{2}}\right)- \sqrt{\frac{2}{\pi}}\,
\exp\left(-\frac{y^2}{2}\right)\right)+\frac{55}{96y}\,
{\rm erfc}\left(\sqrt{2}\,y\right)\nonumber\\
+ \frac{1}{4}\left(\frac{229}{12}-C_E-\lne{1152}\right)
\left(2y\,{\rm erfc}\left(\sqrt{2}\,y\right)-
\sqrt{\frac{2}{\pi}}\,\exp\left(-2y^2\right)\right)
+\frac{y}{8}\,{\cal IL}_{\tau\to\frac{1}{2y^2}}
\left(\frac{\ln{\tau}}{\tau^{3/2}}\,\e^{-2\sqrt{\tau}}\right).\label{thetaos}
\eea The Figure \ref{fig:theta_gamma_os} presents the depletion
interaction potential $\Theta(y)$ and the force $\Gamma(y)$. It
clearly indicates that in comparison to ideal chains the depletion
effect is stronger in the regime of wide slits.\hfill\smallskip\\
{\bf Narrow slits $(y'\ll1)$:}\hfill\\[0.1cm]
Following again the thermodynamic argument, $\omega$ is set to zero
and only bulk and surface contributions are taken into account in
(\ref{decompgrandcan}). One gets: \be \Theta(y) {\approx}\,y\,-
\,\left(1+\frac{2\,\ln\frac{4}{3}\,-\,\frac{1}{2}}{4}\right)\sqrt{\frac{2}{\pi}}\,,\ee
which is also slightly below the depletion potential in comparison
to the case of ideal chains (see Figure \ref{fig:theta_gamma_os}).
The depletion force is unity.\\
 Both approximations for wide, as well as for
narrow slits suggest the depletion effect to be stronger in the case
of excluded volume interactions than for ideal polymer chains (see
Figure 2).

\subsection{Two inert walls}

\begin{figure}[ht!]
\begin{center}
\includegraphics[width=8cm]{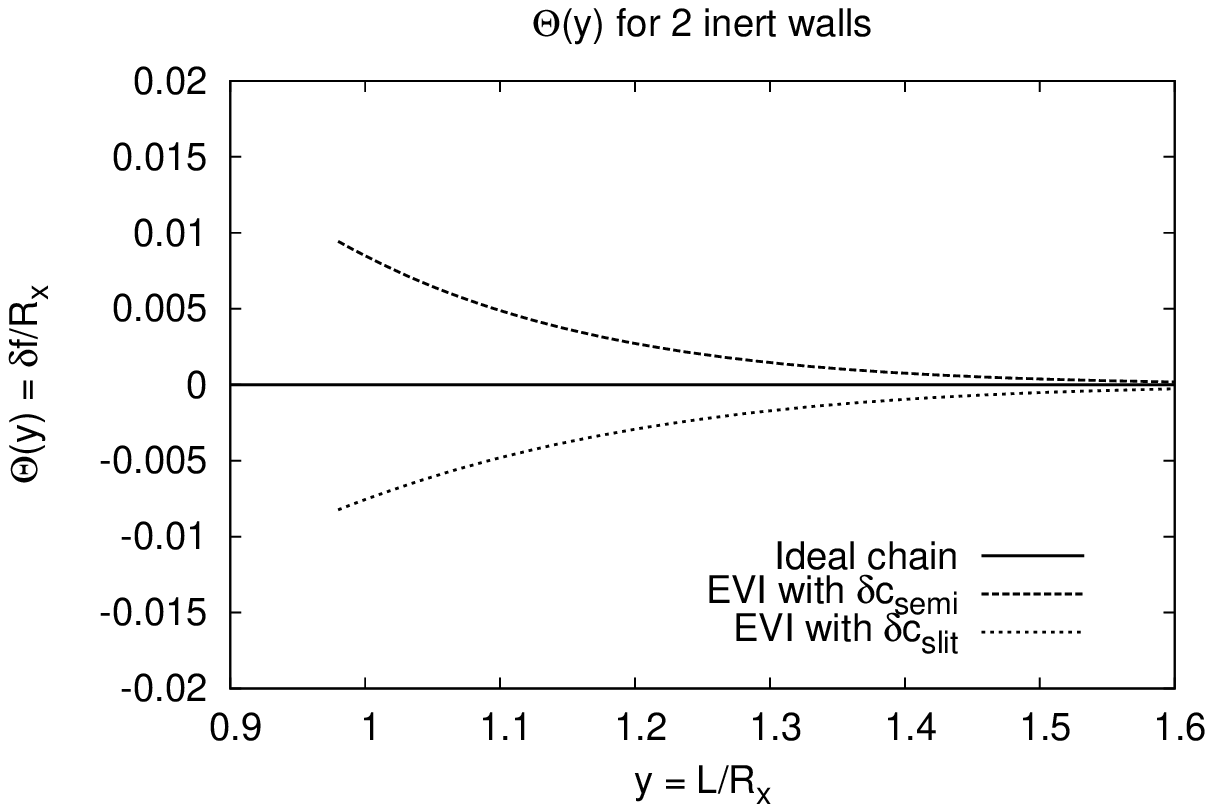}\hspace*{0.3cm}
\includegraphics[width=8cm]{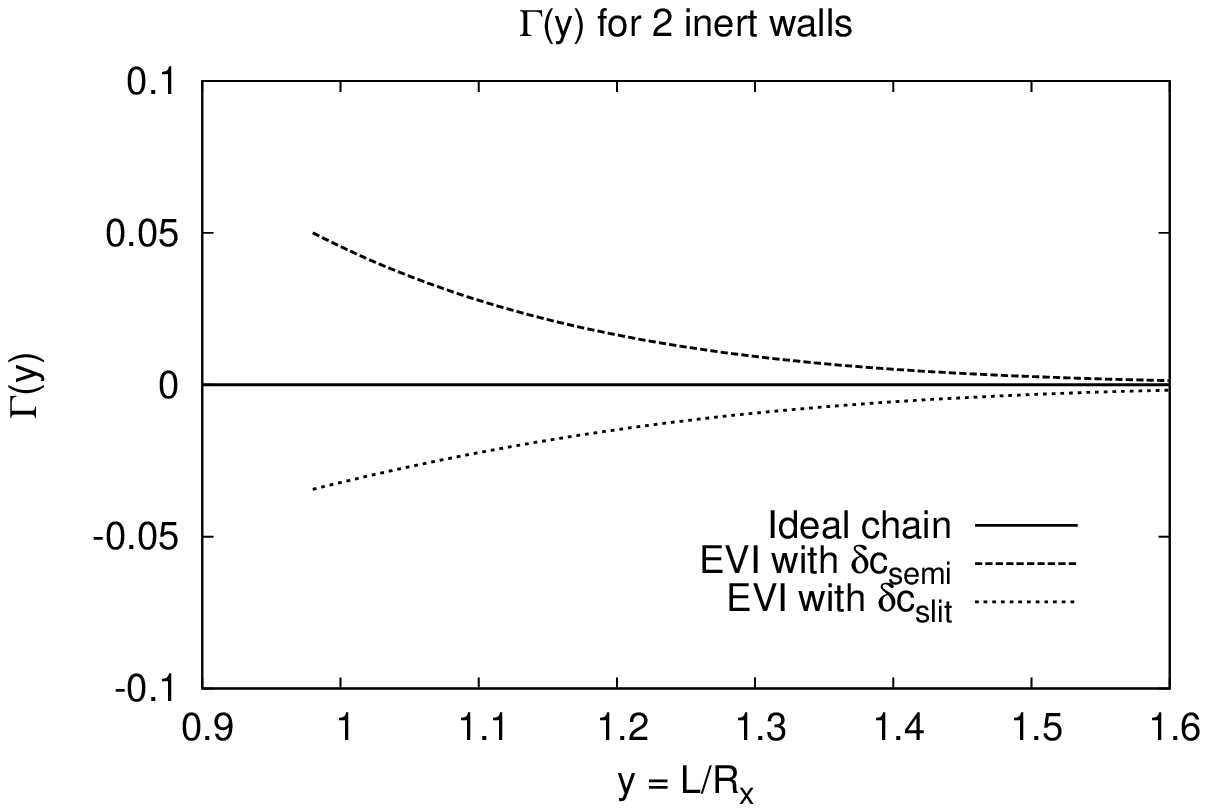}
\caption{The functions $\Theta(y)$ and $\Gamma(y)$ for two inert
walls with and without excluded volume interactions (EVI). Here we
introduced notations: $\delta c_{semi}=\delta c_{i}$ and $\delta
c^{S-S}_{i}=\delta c_{slit}$ with $i=1,2$}
\label{fig:theta_gamma_ss}
\end{center}
\end{figure}

In order to obtain the renormalized total susceptibility for a
system confined by two parallel inert walls we have to apply the
surface renormalization scheme suggested by \cite{DSh98} for both
surfaces at their surface critical point $\c{i}^{sp}$. Starting from
(\ref{1Loop}) we obtain for the renormalized total susceptibility:
\be {\cal
X}_{||\,ren}^{SS}\,L=\frac{L}{\mu^2}\,-\,\frac{1}{2\mu^3}\,
\left(\ln2\,-\,\frac{1}{2}\,-\,\ln\left(1-\e^{-2\mu
L}\right)\right)\,.\label{suscss}\ee The surface contribution has
already been presented in (\ref{surfS}). Lets consider the
asymptotic expansion for wide slits $\mu L>>1$. Taking into account
the surface (\ref{surfS}) and the bulk contributions, the result for
the depletion interaction potential becomes: \be
\Theta(y){\approx}\frac{1}{\sqrt{2\pi}}\,\e^{-2y^2}\,- \,y\,{\rm
erfc}\left(\sqrt{2}\,y\right).\label{thetassdcsemi} \ee
This function and its derivative for the force are plotted in Figure \ref{fig:theta_gamma_ss}.\\

It is obvious that only the wide slit approximation can be applied
here since the usual argument for the narrow slit approximation is
no more valid and $\omega$ does not necessarily vanish.\\
Interestingly, the depletion force turns out to be positive and the
walls are repelled from each other. This means that the chains
rather like to stay in between the slit than leave it. This in turn
means that the chains gain enough energy from attractive
interactions on the walls, which forces them to exert their loss of
entropy (due to the confinement) onto the walls instead
of leaving the slit.\\
It is very instructive to have a more general look on the terms
appearing in the free energy of interaction. If now we take into
account the new normalization conditions for surface-enhancement
constants for slit geometry (see Eqs.(\ref{ccondslit})-
(\ref{deltass})), which assume that we have big, but finite wall
separation $L$, the $\delta f$ can be written as: \be \delta
f=2\,\,{\cal{IL}}_{\mu^2\to \frac{R_x^2}{2}} \left(\frac{\delta
c_{i}^{S-S}-\delta c_{i}}{\mu^4}\right) \,-\,{\cal{IL}}_{\mu^2\to
\frac{R_x^2}{2}} \left(\frac{\ln\left(1-\e^{-2\mu
L}\right)}{2\mu^3}\right)\,. \label{dfssdcsemi}\ee Here $\delta
c^{S-S}$ is the surface-enhancement constant shifts for slit
geometry which appears in the case of finite walls separation and
$\delta c_{i}$ is surface-enhancement constant shift in the case of
infinite walls separation. In the presented approach the same
renormalization of critical values $c_0^{sp}$ was used and equally
the same shifts to the renormalized values were obtained. So the
first term on the r.h.s. just disappeared on the assumption that the
surface-enhancement constant shift on one surface is not affected by
the presence of the second one.\\
In fact this assumption could be doubted and an additional shift
through the influence of the second wall (coupling effect between
the two walls) may appear. Since the interaction potential itself is
purely local, such a coupling effect can only be mediated through
the chain conformations. As a result the number density of monomers
near the walls might differ in comparison to a semi-infinite
constraint and also the shift of the critical point (due to excluded
volume interactions) can change. As already proposed in \cite{GrD08}
this in turn would require a different renormalization scheme for
the surface critical point, where this coupling effect is to be
taken into account. The results of calculations for a slightly
modified surface renormalization scheme which takes into account the
finite surface separation $L$ are introduced at the Appendix B and
are presented in Figure \ref{fig:theta_gamma_ss} as well.

\section{Comparison to Previous Work}
\subsection{Theoretical approach}

\begin{figure}[ht!]
\begin{center}
\includegraphics[width=8cm]{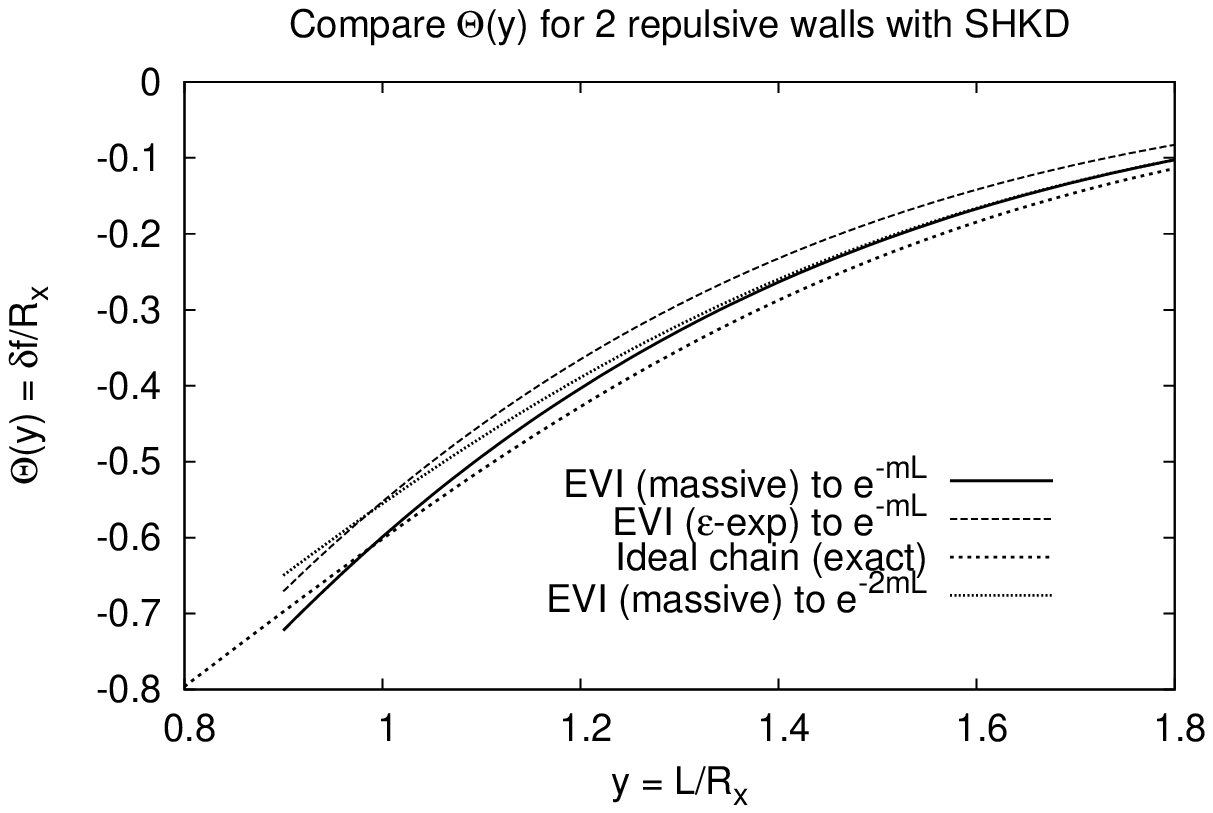}\hspace*{0.3cm}
\includegraphics[width=8cm]{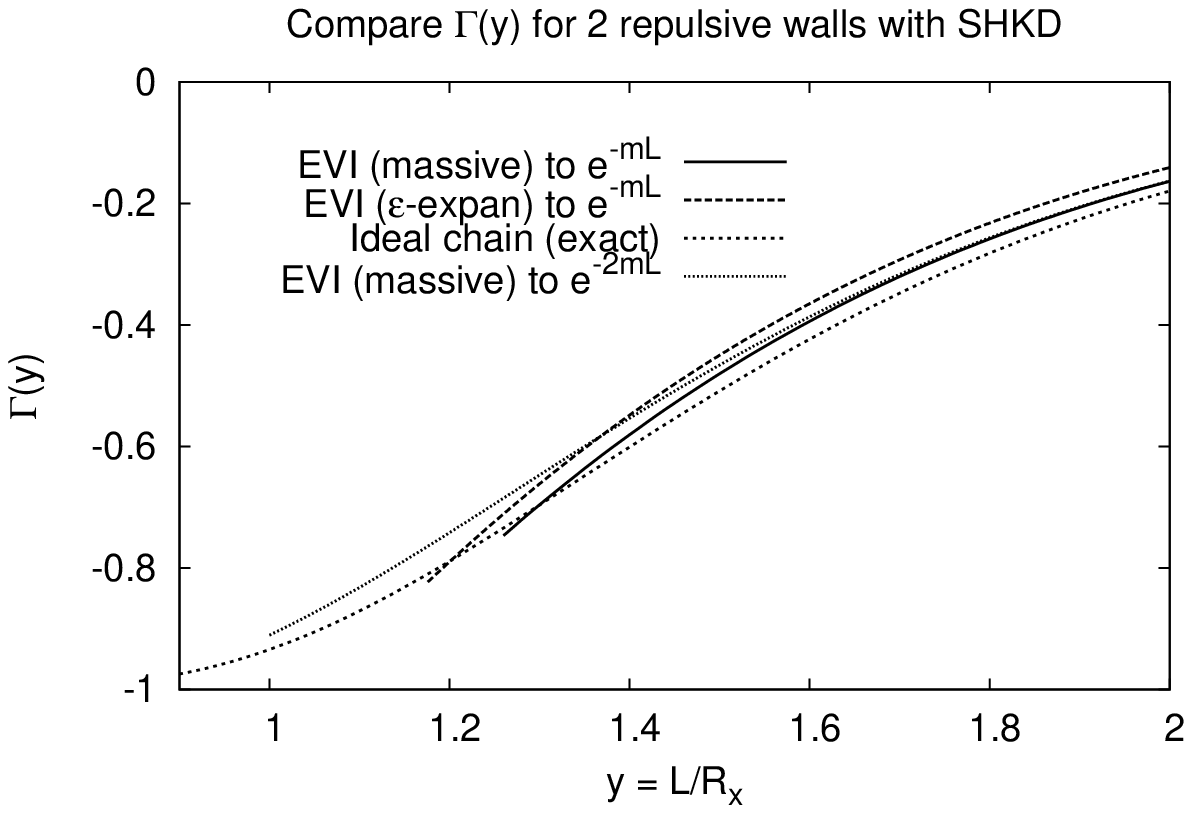}
\caption{The functions $\Theta(y)$ and $\Gamma(y)$ for two repulsive
walls in comparison to \cite{SHKD01}} \label{fig:compshkd}
\end{center}
\end{figure}

As was mentioned in the Introduction, the remarkable progress in the
investigation of the influence of EVI on the depletion interaction
and depletion force between two repulsive walls was achieved by
\cite{E97,SHKD01} via using of dimensionally regularized continuum
version of the field theory with minimal subtraction of poles in
$\epsilon=4-d$, where $d$ is dimensionality of the space. Figure
\ref{fig:compshkd} presents comparison of our results obtained in
the framework of massive field theory at fixed dimensions $d=3$ for
the case of two repulsive walls and results obtained in
\cite{SHKD01}.

The results obtained in the framework of both analytical methods are
in quantitative agreement. But, one notes that the reduction of the
depletion effect due to excluded volume interactions is less
stronger within the massive field approach as compared to an
$\epsilon$-expansion in one-loop order. It should be noted, that we
extended our results up to the next $e^{-2\mu L}$ order. This
allowed us to obtain good matching with approximating results in
narrow slit limit (see Figures 1,2 for $\theta(y)$).

\subsection{Simulations}

One of the possibilities to test reliability of the obtained
analytical results is to compare them to results obtained by Monte
Carlo simulations. In this section we compare our results with
results of MC calculations obtained by \cite{MB98} and \cite{HG04}
for a single polymer chain trapped inside a slit of two repulsive
walls, what corresponds to a canonical ensemble. The canonical free
energy can be obtained via a Legendre transform from the grand
canonical one in the thermodynamical limit ($N,V\to\infty$) in the
form:
 \be
F(N_{l})=\Omega[\mu(N_{l})]+\mu(N_{l})N_{l},\ee with $\Omega$ from
(\ref{tdlimitgrandcan}).

 Thus the reduced canonical force for a one polymer chain
$N_{l}=1$ can be written as its dimensionless counterpart: \be
\frac{K\,L}{k_b\,T}\,=\,\frac{1}{\omega}\frac{d}{d\,L}\big(L\,\omega\big).\label{trapforce}
\ee

It should be mentioned, that both Monte Carlo algorithms (see
\cite{MB98} and \cite{HG04}) differ very much from each other in the
range of analyzed slit widths and chain lengths of the simulated
polymers. In \cite{MB98} an off-lattice bead and spring model for
the self-avoiding polymer chain in $d=3$ dimensions trapped between
two parallel repulsive walls at distance $D$ has been studied by
Monte Carlo methods, using chain lengths up to $N\leq512$ (number of
monomers in the chain) and distances $D$ from $4$ to $32$ (in units
of the maximum spring extension). It was stated that the total force
$K$ exerted on the walls is repulsive and diverges for the case of
narrow slit as \be \frac{K\,L}{k_b\,T}\sim
(\frac{L}{R_{g}})^{-1/\nu},\ee where $R_{g}$ is the radius of
gyration of the polymer chain in unrestricted geometry.

In Ref. \cite{HG04} a lattice Monte Carlo algorithm on a regular
cubic lattice in three dimensions, with $D$ lattice units in
$z$-direction and impenetrable boundaries was applied. The other
directions obeyed periodic boundary conditions. The proposed MC
simulations \cite{HG04} are based on the analytical result obtained
by \cite{E97} for the scaling behavior of partition function for a
chain confined to a slit geometry of width $D$: \eq{
\label{slitlattice}
Z_N(D)\,\propto\,\left(\mu_{\infty}+aD^{-\frac{1}{\nu}}\right)^{-N}\,
\,N^{\gamma_2-1}\,\,D^{\frac{(\gamma_2-\gamma_3)}{\nu}}\,\,, } for
$N,D\to\infty$, but $D\ll N^{\nu}$, where $\mu_{\infty}$ is the
critical fugacity per monomer and $\gamma_d$ is the universal
exponent (see (\ref{RZ})) dependent on the space dimension $d$ and
the parameter $a$ is a universal amplitude.  The critical fugacity
means the averaged inverse number of possible steps at each site. In
\cite{HG04} universal amplitudes and exponents for the partition
function of a chain trapped in the slit with respect to that of a
free chain have been obtained through analyzing the statistics for
different $D$ and number of chain monomers up to $N\leq80\,000$.
Also both cases of ideal chains (modeled as a simple random walk
(RW)) and chains with excluded volume interactions (modeled via
self-avoiding walks (SAW)) have been studied. In the case of an
ordinary random walk on a regular cubic lattice in three dimensions
one has obviously $\mu_{\infty}=\frac{1}{6}$ and $\gamma_{d=3}=1$.
In the case of a SAW on such a lattice it is clear that at least
$\mu_{\infty}\geq\frac{1}{5}$. From Eq. (\ref{slitlattice}) one may
obtain the force exerted onto the walls in units of $k_B\,T$ as:
\eq{ \tilde{K}\,=k_B\,T\,\frac{d}{dD}\,\ln\big(Z_N(D)\big)\,. } In
the limit ($D\ll N^{\nu}$, $N,D\to\infty$) $\tilde{K}$ becomes: \eq{
\tilde{K}\,=
\,k_B\,T\left(\frac{Na}{\nu\mu_{\infty}}\,D^{-1-\frac{1}{\nu}}\right)\,.
} One should note that all functions here are in terms of
dimensionless length scales, the number of lattice sites ($D$ and
$N$). In order to compare with  our results it must be translated
into terms of $L$ and $R_g$. Apparently $L=uD$, with $u$ denoting
the lattice spacing, and the reduced, dimensionless force reads:
\eq{k=\frac{\tilde{K}\,D}{k_B\,T}\,=
\,\frac{a}{\nu_3\,\mu_{\infty}}\,\left(\frac{6}{\chi^2_d}\right)^{\frac{1}{2\nu_3}}\,
\left(\frac{L}{R_g}\right)^{-\frac{1}{\nu_3}}\,, } where we
 take
into account the general relation (e.g. \cite{CJ90}): \eq{
R_g^2=\chi^2_3\frac{b^2\,N^{2\nu_3}}{6} } in $d=3$ dimensions.
Parameter $b$ denotes the (effective) segment length of the polymer
model under consideration. In the case of RW and SAW on a cubic
lattice one has simply $b=u$ because the segment length in these
models is naturally provided by the lattice parameter $u$. In
\cite{HG04} the universal amplitude $a$ for the case of ideal chains
\begin{figure}[ht!]
\begin{center}
\includegraphics[width=10cm]{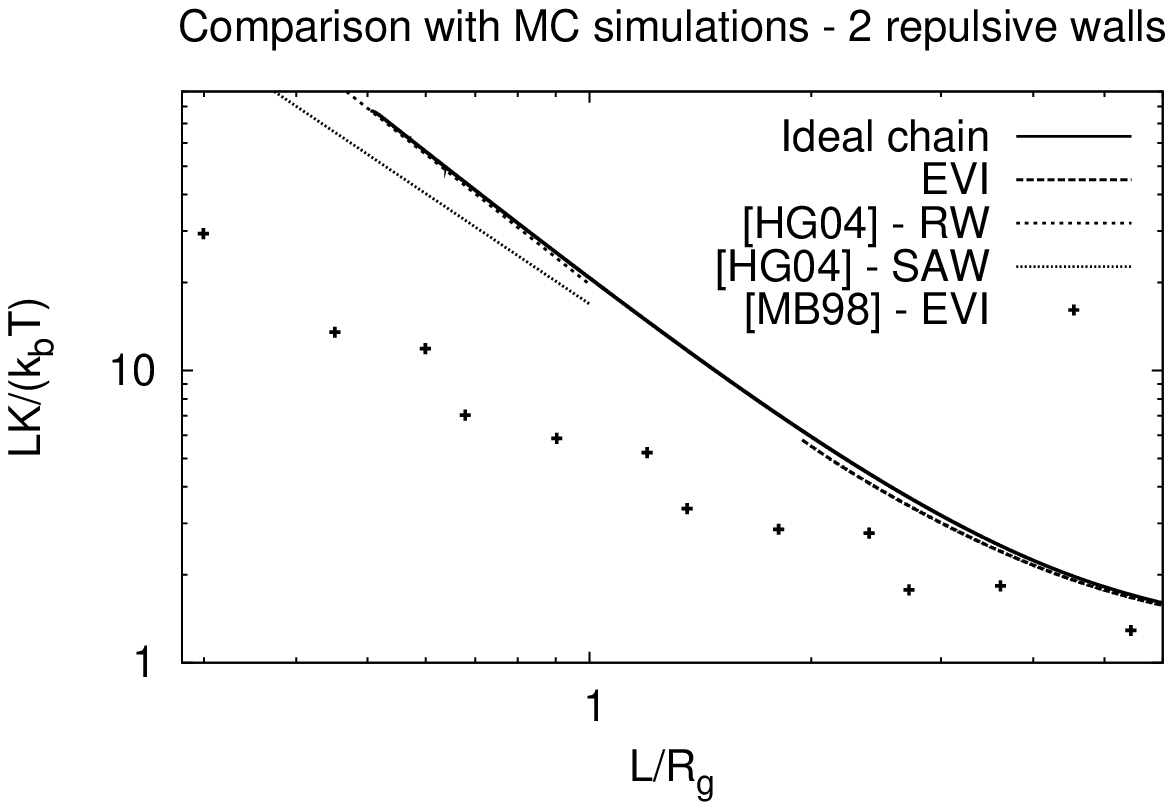}
\caption{Comparison of our theoretical results with Monte Carlo
simulations for a {\it trapped} chain between two repulsive walls}
\vspace*{0.2cm}\centering {\small The plots Ideal chain (exact) and
EVI (wide slit) represent the results of our calculations. MC - RW
and MC - SAW are due to the estimated asymptotic behaviour in the
narrow slit limit by \cite{HG04} for random walks and self avoiding
walks. MC - EVI are the results obtained by \cite{MB98}. }
\label{fig:comp_MC}
\end{center}
\end{figure}
was found as $a\approx0.2741$, which is very close to the exact
value, computed analytically in \cite{E97}, of $a=\frac{\pi^2}{36}$.
Taking into account that for ideal chain $\nu=0.5$, $\chi_d=1$ and
$\mu_{\infty}=\frac{1}{6}$ the force becomes: \eq{
\label{idealcanforce}
k_{id}\,=\,2\,\pi^2\,\left(\frac{L}{R_g}\right)^{-2}\,\,. } In
Figure \ref{fig:comp_MC} this asymptotic behaviour for narrow slits
 is clearly recovered by our results for ideal chains, where the
narrow slit limit is valid. By contrast, for a SAW in Ref.
\cite{HG04} was suggested the value $a\approx0.448$. Taking into
account the values for $\nu\approx0.588$, $\chi_3\approx0.958$
\cite{CJ90} and $\mu_{\infty}\approx0.2135$ the reduced force can be
written as \eq{ \label{evicanforce}
k_{saw}\,\approx\,16.95\,\left(\frac{L}{R_g}\right)^{-1.7}\,\,. }
The result Eq.(\ref{evicanforce}) is presented in Figure
\ref{fig:comp_MC} in its regime of validity and compared to our
theoretical results for a trapped chain with EVI, which are valid
for the wide slit regime. As it easily can be see from Figure
\ref{fig:comp_MC}, the result (\ref{evicanforce}) very well fit to
our predictions in wide slit limit. Also, in Figure
\ref{fig:comp_MC} the results obtained by the authors of Ref.\
\cite{MB98} are plotted and one notes a qualitative agreement to our
predictions. One of the possible reason for the remaining deviations
with results of Ref. \cite{MB98} is that the chain in the MC
simulation is too short in order to compare with results of
field-theoretical calculations. It should be noted, that at the
moment no simulations concerning two inert walls or one inert/ one
repulsive wall exist.
\subsection{Experiment}

In Ref.\cite{RBL98} an experimental study of the depletion effect
between a spherical colloidal particle immersed in a dilute solution
of nonionic linear polymer chains and a wall of the container
through total internal reflection microscopy was analyzed. Using the
Derjaguin \cite{D34} approximation we could compare our theoretical
results with experimental data in the case when the radius of the
spherical colloid particle $R$ is much larger than radius of
gyration $R_{g}$ and the closest distance $a$ between particle and
the surface. The deviation of the experimental setup to the
presented theoretical approach connected with the fact that the {\it
second} wall is not plane but curved. Summing up the depletion
potential per volume unit for the case of two plane surfaces in the
margins of the curved volume allows to estimate the depletion
effects in the case of a sphere and a wall. In the experiment by
\cite{RBL98} the radius of gyration was measured as
$R_g=0.101\,\,\mu m$ and the colloidal particle was reported to have
a radius $R=1.5\,\,\mu m$. Straightforward application of the
Derjaguin \cite{D34} approximation yields: \be
\frac{\phi_{depl}(a)}{n_p\,k_b\,T}\,=
\,2\,\pi\,R_x^2\int\limits_{\frac{a}{R_x}}^{\frac{a+R}{R_x}}
dy\,\big(R+a-R_xy\big)\,\Theta\big(y\big), \label{derjaguin} \ee
with $a$ the minimal distance between the sphere and the wall. Since
in the range of $y$ the last two terms in the parenthesis are much
smaller in comparison to the first one we can assume that: \be
\frac{\phi_{depl}(a)}{n_p\,k_b\,T}\,\approx
\,2\,\pi\,R\,R_x^2\int\limits_{\frac{a}{R_x}}^{\infty}
dy\,\Theta\big(y\big)~. \ee The experimental data in comparison to
our theoretical prediction are plotted
\begin{figure}[ht!]
\begin{center}
\includegraphics[width=9cm]{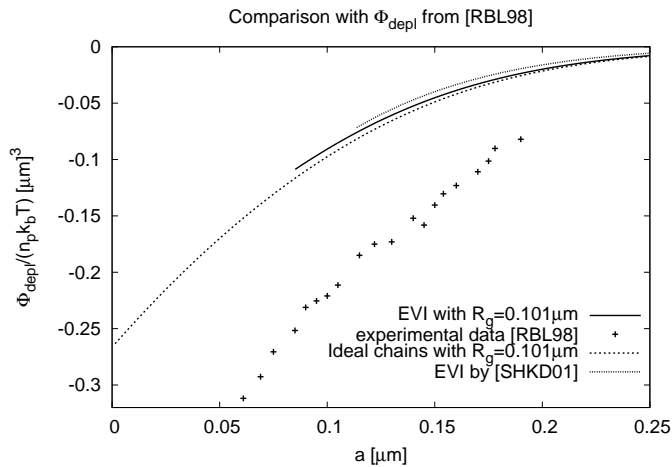}
\caption{Comparison of approximated theoretical results with
experimental observation, due to \cite{RBL98}} \label{fig:compexp}
\end{center}
\end{figure}
in Figure \ref{fig:compexp}. It should be mentioned, that our
results obtained in the framework of the massive field theory are
situated slightly closer to the experimental data than previous
theoretical results obtained in the framework of the dimensionally
regularized continuum version of the field theory with minimal
subtraction of poles in $\epsilon=4-d$ \cite{SHKD01}. Unfortunately,
this shift is not enough in order to obtain quantitative agreement
with experimental data. But, the obtained theoretical curves in
Figure \ref{fig:compexp} are in qualitative agreement with
experimental data. The quantitative discrepancy can be removed if we
use the radius of gyration as adjusting parameter by analogy as it
was done in \cite{SHKD01}. From another side this indicate about
importance of further theoretical investigations of depletion
interaction potential and depletion force in the crossover region
from wide to narrow slit.
\section{Conclusions}
Using the massive field theory approach directly at fixed dimensions
$d=3$ we calculated the depletion interaction potential and
depletion force between two repulsive, two inert and one repulsive
and one inert walls confining a dilute solution of long flexible
polymer chains. The obtained calculations for all cases of
polymer-surface interactions were performed for the ideal chain and
real polymer chain with excluded volume interactions in the wide
slit regime. Besides, we used some assumptions which allowed us to
estimate the depletion interaction potential in the region of narrow
slit. Our results are obtained up to the next $e^{-2\mu L}$ order in
comparison with results of $\epsilon$-expansion \cite{SHKD01}. Our
investigations include modification of renormalization scheme for
the case of two inert walls (or mixed walls) situated on big, but
finite distance $L$ with $L\gtrsim R_{g}$ such that the polymer
chain is still not deformed too much from its original size in the
bulk. The obtained results indicate that the reduction of the
depletion effect due to excluded volume interactions is less
stronger within the massive field theory approach as compared to the
dimensionally regularized continuum version of the field theory with
minimal subtraction of poles in $\epsilon=4-d$ \cite{SHKD01} in
one-loop order. We found very good agreement with Monte Carlo
simulation data \cite{HG04} and \cite{MB98} for the case of two
repulsive walls. Taking into account Derjaguin approximation we
obtained good qualitative agreement with experimental data
\cite{RBL98} for the depletion potential between a spherical
colloidal particle of big radius and repulsive wall. From comparison
of obtained theoretical results and experimental data we can see
that the results obtained in the framework of the massive field
theory are situated slightly closer to experimental data. But, this
shift is not enough in order to obtain good quantitative agreement
with experiment. Interesting fact is that even the taking into
account the excluded volume interaction between the monomers of the
polymer chain do not resolve completely this problem. One of the
possible ways to find a good agreement could be connected with
further theoretical investigation of crossover region from wide to
narrow slit.
\section*{Acknowledgments}
We gratefully acknowledge fruitful discussions with H.W. Diehl.
 This work in part was supported by grant from the Alexander von Humboldt Foundation
(Z.U.).

\renewcommand{\theequation}{A1.\arabic{equation}}
\section*{Appendix A: The surface contributions}
\setcounter{equation}{0}

To calculate the function $\Upsilon$ defined in  Eq.(\ref{dfcrit}),
we need the free propagator for a semi-infinite system confined by a
surface at $z=0$. This free full propagator has a form \cite{DSh98}:
\be {\tilde G}^*_{ij\,HS}({\bf p},{\bf p'},z,z';\mu_0,c_0)=
(2\pi)^{d-1}\delta_{ij}\delta({\bf p}+{\bf p'}) {\tilde G}_{HS}({\bf
p},z,z',\mu_0,c_0)~, \label{semipropagator} \ee with \ali{
\label{g0semi} {\tilde G}_{HS}({\bf
p},z,z',\mu_0,c_0):\,&=\frac{1}{2\k_0}
\left(\e^{-\k_0|z-z'|}+\frac{\k_0-c_0}{\k_0+c_0}\e^{-\k_0(z+z')}\right)~,\\[0.2cm]
&\hspace*{4cm}{\rm where}~~\k_0=\sqrt{p^2+\mu_0^2}~.\nonumber } In
the zero-loop order we have: \be
\Upsilon_{i}=\frac{1}{\mu^3}\,\frac{c_i}{\mu+c_i}\,.
\label{0loopsemi}\ee In one-loop order the calculation for Dirichlet
boundary conditions on the surface (or $\frac{c}{m}\to\infty$)
yields after renormalization in fixed dimensions $d=3$: \be
\Upsilon^{D}=\frac{1}{\mu^3}\,\left(1-\frac{n+2}{n+8}\,\tilde{v}\,\ln\frac{9}{8}\right)\,.\label{1loopsemiD}\ee
And for Neumann boundary conditions ($c=0$) after renormalization we
obtain: \be
\Upsilon^{N}=\frac{\tilde{v}}{\mu^3}\,\left(\ln2-\frac{1}{2}\right)\frac{n+2}{n+8}\,,
\label{1loopsemiN}\ee where we introduced rescaled renormalized
coupling constant $\tilde{v}$ in the form:
$\tilde{v}=\frac{(n+8)}{6}\frac{\Gamma({\epsilon/2})}{(4\pi)^{d/2}}
v$. The correspondent fixed point in one-loop order approximation is
$\tilde{v}^{*}=1$.

\renewcommand{\theequation}{A2.\arabic{equation}}
\section*{Appendix B: The case of finite slit separation for two inert walls}
\setcounter{equation}{0}

Taking into account the new $\delta c_i^{S-S}$ (see
Eq.(\ref{cslitss}) we can calculate $\delta f$ in accordance with
Eq.(\ref{dfssdcsemi}) for the case of big, but finite slit
separation $L$. We obtain:  \be \delta
f\,{\approx}\,\,{\cal{IL}}_{\mu^2\to
\frac{R_x^2}{2}}\left(\frac{~~2\,\,\Delta^{(S-S)}}{\mu^4}+\frac{\e^{-2\mu
L}}{2\mu^3}\right)\,. \ee After substitution of $\Delta^{(S-S)}$
from (\ref{deltass}) the result for $\delta f$ is: \be \delta
f\,{\approx}-\,{\cal{IL}}_{\mu^2\to
\frac{R_x^2}{2}}\left\lbrace\left( \frac{1}{\mu L}+C_E+7\ln2\,-6+\ln
\mu L\,-\e^{4\mu L}\,{\rm Ei}(-4\mu L)\right)\frac{\e^{-2\mu
L}}{2\mu^3}\right\rbrace. \ee  If we carry out the inverse Laplace
transform, the result for $\Theta(y)$ in the wide slit limit is:
\mli{ \Theta(y)\,{\approx}\,-\frac{C_E\,+\,7\ln2\,-7}{2}
\left(\sqrt{\frac{2}{\pi}}\,\e^{-2y^2}\,-\,2\,y\,{\rm
erfc}\left(\sqrt{2}\,\,y\right)\right)\, -\,\frac{1}{4\,y}\,{\rm
erfc}\left(\sqrt{2}\,\,y\right)\\-\,\frac{y}{4}\,\,{\cal{IL}}_{\tau\to
\frac{1}{2y^2}}\left(\frac{\e^{-2\sqrt{\tau}}}{\tau^{3/2}}\ln\tau\right)
\,+\,\frac{y}{2}\,\,{\cal{IL}}_{\tau\to
\frac{1}{2y^2}}\left(\frac{\e^{2\sqrt{\tau}}}{\tau^{3/2}}{\rm
Ei}\left(-4\sqrt{\tau}\right)\right)\,. } In contrast to
(\ref{thetassdcsemi}) this expression is indeed negative. Thus, if
we perform calculations for the depletion interaction potential and
the depletion force including big, but finite slit separation $L$ we
obtain, that force for the case of two inert walls change character
and becomes attractive. In Figure \ref{fig:theta_gamma_ss} the
depletion interaction potential and the depletion force obtained in
the framework of this alternative renormalization  scheme  with
$\delta c_{slit}$ are plotted in comparison to the results obtained
via the original renormalization using $\delta c_{semi}$. Here we
introduced for convenience the following notations: $\delta
c_{semi}=\delta c_{i}$ and $\delta c^{S-S}_{i}=\delta c_{slit}$ with
$i=1,2$.

\end{document}